%% file: sample-acmtog-SIGGRAPH-submission.tex
\documentclass[acmtog, authorversion]{acmart}

\usepackage{booktabs} 

\usepackage{ amssymb }
\usepackage{wrapfig}
\usepackage{graphicx}
\usepackage{subfig}
\usepackage{multirow}
\usepackage{algorithm}
\usepackage[noend]{algpseudocode}

\acmJournal{TOG}

\setcopyright{acmlicensed}
\acmJournal{TOG}
\acmYear{2020}
\acmVolume{39}
\acmNumber{4}
\acmArticle{80}
\acmMonth{7}
\acmDOI{10.1145/3386569.3392403}

\newcommand\mycolor[1]{\textcolor{black}{#1}}

\begin{document}
\title{Single Image HDR Reconstruction Using a CNN with Masked Features and Perceptual Loss}

\author{Marcel Santana Santos}
\affiliation{%
 \institution{Centro de Inform\'atica, Universidade Federal de Pernambuco}}
\email{mss8@cin.ufpe.br}

\author{Tsang Ing Ren}
\affiliation{%
 \institution{Centro de Inform\'atica, Universidade Federal de Pernambuco}}
\email{tir@cin.ufpe.br}

\author{Nima Khademi Kalantari}
\affiliation{%
\institution{Texas A\&M University}}
\email{nimak@tamu.edu}

\begin{abstract}
Digital cameras can only capture a limited range of real-world scenes' luminance, producing images with saturated pixels. \mycolor{Existing} single image high dynamic range (HDR) reconstruction methods attempt to expand the range of luminance, but are not able to hallucinate plausible textures, producing results \mycolor{with artifacts} in the saturated areas. In this paper, we present a novel learning-based approach to reconstruct an HDR image by recovering the saturated pixels of an input LDR image in a visually pleasing way. Previous deep learning-based methods apply the same convolutional filters on well-exposed and saturated pixels, creating ambiguity during training and leading to checkerboard and halo artifacts. To overcome this problem, we propose a feature masking mechanism that reduces the contribution of the features from the saturated areas. Moreover, we adapt the VGG-based perceptual loss function to our application to be able to synthesize visually pleasing textures. Since the number of HDR images for training is limited, we propose to train our system in two stages. Specifically, we first train our system on a large number of images for image inpainting task and then fine-tune it on HDR reconstruction. Since most of the HDR examples contain smooth regions that are simple to reconstruct, we propose a sampling strategy to select challenging training patches during the HDR fine-tuning stage. We demonstrate through experimental results that our approach can reconstruct visually pleasing HDR results, better than the current state of the art on a wide range of scenes.
\end{abstract}

%
%


%
%


\begin{teaserfigure}
\centering
\includegraphics[width=7.0in]{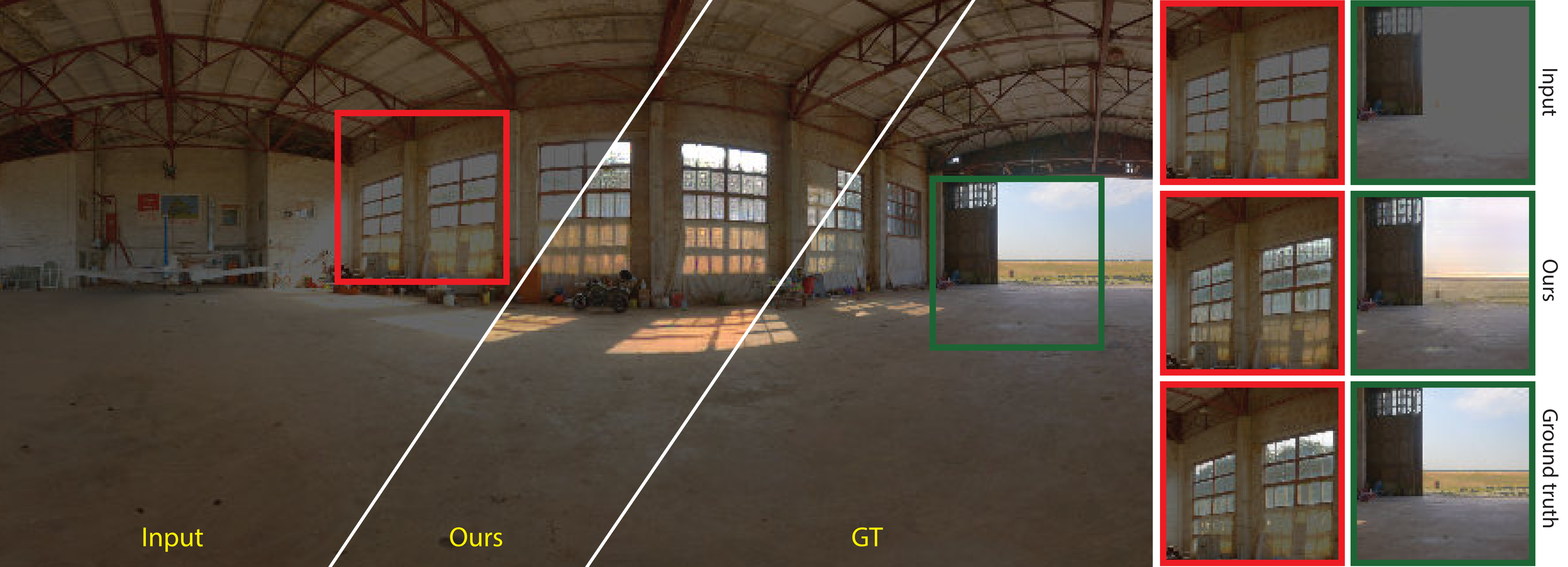}
\vspace{-7pt}
\caption{We propose a novel deep learning system for single image HDR reconstruction by synthesizing visually pleasing details in the saturated areas. We introduce a new feature masking approach that reduces the contribution of the features computed on the saturated areas, to mitigate halo and checkerboard artifacts. To synthesize visually pleasing textures in the saturated regions, we adapt the VGG-based perceptual loss function to the HDR reconstruction application. Furthermore, to effectively train our network on limited HDR training data, we propose to pre-train the network on inpainting task. Our method can reconstruct regions with high luminance, such as the bright highlights of the windows (red inset), and generate visually pleasing textures (green insert). See Figure~\ref{fig:comparison} for comparison against several other approaches. All images have been gamma corrected for display purposes.}
\label{fig:teaser}
\end{teaserfigure}

\maketitle

\input{samplebody-journals}

\end{document}

%% file: samplebody-journals.tex
\vspace{40pt}
\section{Introduction}

The illumination of real-world scenes is high dynamic range, but standard digital cameras sensors can only capture a limited range of luminance. Therefore, these cameras typically produce images with under/over-exposed areas. A large number of approaches propose to generate a high dynamic range (HDR) image by combining a set of low dynamic range images (LDR) of the scene at different exposures \cite{debevec1997recovering}. However, these methods either have to handle the scene motion~\cite{ kalantari2017deep, wu2018deep, kang2003high, sen2012robust, hu2013hdr, oh2014robust} or require specialized bulky and expensive optical systems \cite{mcguire2007optical, tocci2011versatile}. Single image dynamic range expansion approaches avoid these limitations by reconstructing an HDR image using one image. These approaches can work with images captured with any standard camera or even recover the full dynamic range of legacy LDR content. As a result, they have attracted considerable attention in recent years. 

Several existing methods extrapolate the light intensity using heuristic rules \cite{banterle2006inverse, rempel2007ldr2hdr, bist2017tone}, but are not able to properly recover the brightness of saturated areas as they do not utilize context. On the other hand, recent deep learning approaches \cite{endo2017deep, lee2018deep, eilertsen2017hdr} systematically utilize contextual information using convolutional neural networks (CNNs) with large receptive fields. However, these methods usually produce results with blurriness, checkerboard, and halo artifacts in saturated areas.

In this paper, we propose a novel learning-based technique to reconstruct an HDR image by recovering the missing information in the saturated areas of an LDR image. We design our approach based on two main observations. First, applying the same convolutional filters on well-exposed and saturated pixels, as done in previous approaches, results in ambiguity during training and leads to checkerboard and halo artifacts. Second, using simple pixel-wise loss functions, utilized by most existing approaches, the network is unable to hallucinate details in the saturated areas, producing blurry results. To address these limitations, we propose a feature masking mechanism that reduces the contribution of features generated from the saturated content by multiplying them to a soft mask. With this simple strategy, we are able to avoid checkerboard and halo artifacts as the network only relies on the valid information of the input image to produce the HDR image. Moreover, inspired by image inpainting approaches, we leverage the VGG-based perceptual loss function, introduced by ~\citet{gatys2016image}, and adapt it to the HDR reconstruction task. By minimizing our proposed perceptual loss function during training, the network can synthesize visually realistic textures in the saturated areas.

Since a large number of HDR images, required for training a deep neural network, are currently not available, \mycolor{we perform} the training in two stages. In the first stage, we train our system on a large set of images for the inpainting task. During this process, the network leverages a large number of training \mycolor{samples} to learn an internal representation that is suitable for synthesizing visually realistic texture in the incomplete regions. In the next step, we fine-tune this network on the HDR reconstruction task using a set of simulated LDR and their corresponding ground truth HDR images. Since most of the HDR examples contain smooth regions that are simple to reconstruct, we propose a simple method to identify the textured patches and only use them for fine-tuning.

Our approach can reconstruct regions with high luminance and hallucinate textures in the saturated areas, as shown in Figure~\ref{fig:teaser}. We demonstrate that our approach can produce better results than the state-of-the-art methods both on simulated images (Figure~\ref{fig:comparison}) and on images taken with real-world cameras (Figure~\ref{fig:real_world}). In summary, we make the following contributions:

\begin{enumerate}
    \item We propose a feature masking mechanism to avoid relying on the invalid information in the saturated regions (Section~\ref{sec:softpconv}). This masking approach significantly reduces the artifacts and improves the quality of the final results (Figure~\ref{fig:masking_std}).
    
    \item We adapt the VGG-based perceptual loss function to the HDR reconstruction task (Section~ \ref{sec:hybridloss}). Compared to pixel-wise loss functions, our loss can better reconstruct sharp textures in the saturated regions (Figure~\ref{fig:loss_comp}).
    
    \item We \mycolor{propose to pre-train} the network on inpainting before fine-tuning it on HDR generation (Section~ \ref{sec:train}). We demonstrate that the pre-training stage is essential for synthesizing visually pleasing textures in the saturated areas (Figure~\ref{fig:two_stage}).
    
    \item We propose a simple strategy for identifying the textured HDR areas to improve the performance of training (Section~\ref{sec:patch_sampling}). This strategy improves the network ability to reconstruct sharp details (Figure~\ref{fig:two_stage}).
    
\end{enumerate}

\section{Related Work}

The problem of single image HDR reconstruction, also known as inverse tone-mapping \cite{banterle2006inverse}, has been extensively studied in the last couple of decades. However, this problem remains a major challenge as it requires recovering the details from regions with missing content. In this section, we discuss the existing techniques by classifying them into two categories of non-learning and learning methods.

\subsection{Non-learning Methods}

Several approaches propose to perform inverse tone-mapping using global operators. \citet{landis2002production} applies a linear or exponential function to the pixels of the LDR image above a certain threshold. \citet{bist2017tone} approximates tone expansion by a gamma function. They use the characteristics of the human visual system to design the gamma curve. \citet{luzardo2018fully} improve the brightness of the result by utilizing an operator based on the mid-level mapping.

A number of techniques propose to handle this application through local heuristics. \citet{banterle2006inverse} use median-cut \cite{debevec2005median} to find areas with high luminance. They then generate an expand-map to extend the range of luminance in these areas, using an inverse operator. \citet{rempel2007ldr2hdr} also utilize an expand-map but use a Gaussian filter followed by an edge-stopping function to enhance the brightness of saturated areas. \citet{kovaleski2014high} extend the approach by \citet{rempel2007ldr2hdr} using a cross bilateral filter. These approaches simply extrapolate the light intensity by using heuristics and, thus, often fail to recover saturated highlights, introducing unnatural artifacts.

A few approaches propose to handle this application by incorporating user interactions in their system. \citet{didyk2008enhancement} enhance bright luminous objects in video sequences by using a semi-automatic classifier to classify saturated regions as lights, reflections, or diffuse surfaces. \citet{wang2007high} recover the textures in the saturated areas by transferring details from the user-selected regions. Their approach demands user interactions that take several minutes, even for an expert user. In contrast to these methods, we propose a learning-based approach to systematically reconstruct HDR images from a wide range of different scenes, instead of relying on heuristics strategies and user inputs.

\subsection{Learning-based Methods}

In recent years, several approaches have proposed to tackle this application using deep convolutional neural networks (CNN). Given a single input LDR image, \citet{endo2017deep} use an auto-encoder \cite{hinton2006reducing} to generate a set of LDR images with different exposures. These images are then combined to reconstruct the final HDR image. \citet{lee2018deep} chain a set of CNNs to sequentially generate the bracketed LDR images. Later, they propose \cite{alee2018deep} to handle this application through a recursive conditional generative adversarial network (GAN) \cite{goodfellow2014generative} combined with a pixel-wise $l_1$ loss.

In contrast to these approaches, a few methods \cite{eilertsen2017hdr, yang2018image, marnerides2018expandnet} directly reconstruct the HDR image without generating bracketed images. \citet{eilertsen2017hdr} use a network with U-Net architecture to predict the values of the saturated areas, whereas linear non-saturated areas are obtained from the input. \citet{marnerides2018expandnet} present a novel dedicated architecture for end-to-end image expansion. \citet{yang2018image} reconstruct HDR image for image correction application. They train a network for HDR reconstruction to recover the missing details from the input LDR image, and then a second network transfers these details back to the LDR domain.

While these approaches produce state-of-the-art results, \mycolor{their synthesized images often contains halo and checkerboard artifacts and lacks textures} in the saturated areas. This is mainly because of using standard convolutional layers and pixel-wise loss functions. Note that, \mycolor{several recent methods \cite{alee2018deep, xu2019gan, ning2018learning, kim2019jsi} use adversarial loss instead of pixel-wise loss functions, but they still do not demonstrate results with high-quality textures. This is potentially because the problem of HDR reconstruction is constrained in the sense that the synthesized content should properly fit the input image using a soft mask. Unfortunately, GANs are known to have difficulty handling these scenarios \cite{Bau:Ganpaint:2019}. } In contrast, we propose a feature masking strategy and a more constrained VGG-based perceptual loss to effectively train our network and produce results with visually pleasing textures.

\section{APPROACH}
\label{sec:sim}

Our goal is to reconstruct an HDR image from a single LDR image by recovering the missing information in the saturated highlights. \mycolor{We achieve} this using a convolutional neural network (CNN) that takes an LDR image as the input and estimates the missing HDR information in the bright regions. We compute the final HDR image by combining the well-exposed content of the input image and the output of the network in the saturated areas. Formally, we reconstruct the final HDR image $\hat{H}$, as follows:

\vspace{-0.1in}
\begin{equation}
\begin{aligned}
    \hat{H} = M \odot  T^{\gamma} + (1 - M) \odot [\exp(\hat{Y}) - 1],
    \label{eq:H}
\end{aligned}
\end{equation}  

\begin{wrapfigure}{r}{0.21\textwidth}
\vspace{-0.0in}
\includegraphics[width=\linewidth]{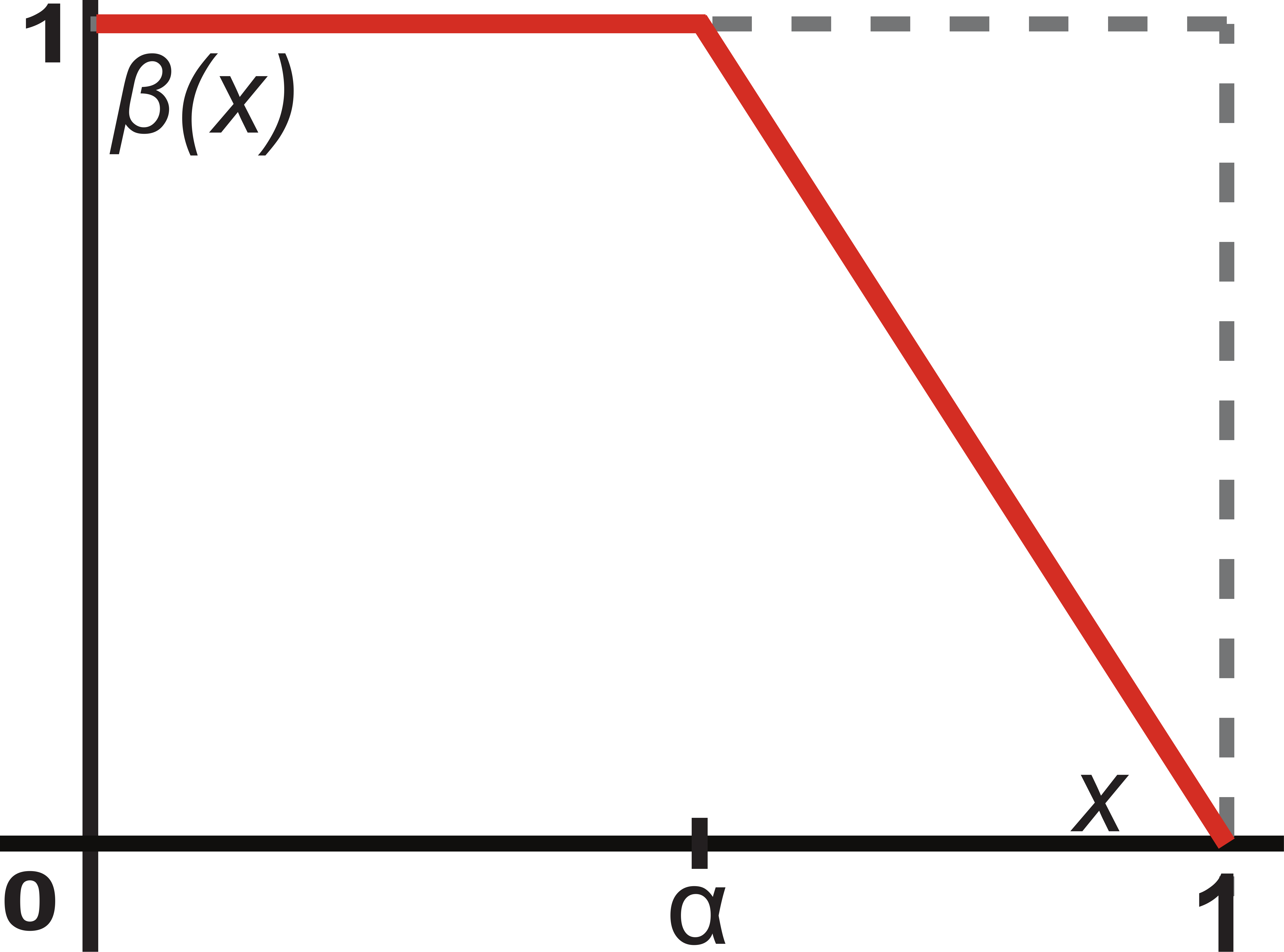}
\vspace{-0.3in}
\caption{We use this function to measure how well-exposed a pixel is. The value 1 indicates that the pixel is well-exposed, while 0 is assigned to the pixels that are fully saturated. In our implementation, we set the threshold $\alpha = 0.96$.}
\vspace{-0.20in}
\label{fig:beta}
\end{wrapfigure}

\noindent where the $ \gamma = 2.0 $ is used to transform the input image to the linear domain, and $\odot$ denotes element-wise multiplication. Here, $ T $ is the input LDR image in the range $ [0, 1] $, $\hat{Y} $ is the network output in the logarithmic domain (Section~\ref{sec:hybridloss}), and $M$ is a soft mask with values in the range $[0, 1]$ that defines how well-exposed each pixel is. We obtain this mask by applying the function $ \beta(\cdot)$ (see Figure~\ref{fig:beta}) to the input image, i.e., $M = \beta(T)$. In the following sections, we discuss our proposed feature masking approach, loss function, as well as the training process.

\subsection{Feature Masking}
\label{sec:softpconv}

Standard convolutional layers apply the same filter to the entire image to extract a set of features. This is reasonable for a wide range of applications, such as image super-resolution \cite{dong2015image}, style transfer \cite{gatys2016image}, and image colorization \cite{zhang2016colorful}, where the entire image contains valid information. However, in our problem, the input LDR image contains invalid information in the saturated areas. Since meaningful features cannot be extracted from the saturated contents, na\"ive application of standard convolution introduces ambiguity during training and leads to visible artifacts (Figure~\ref{fig:masking_std}).

\begin{figure}
  \includegraphics[width=\linewidth]{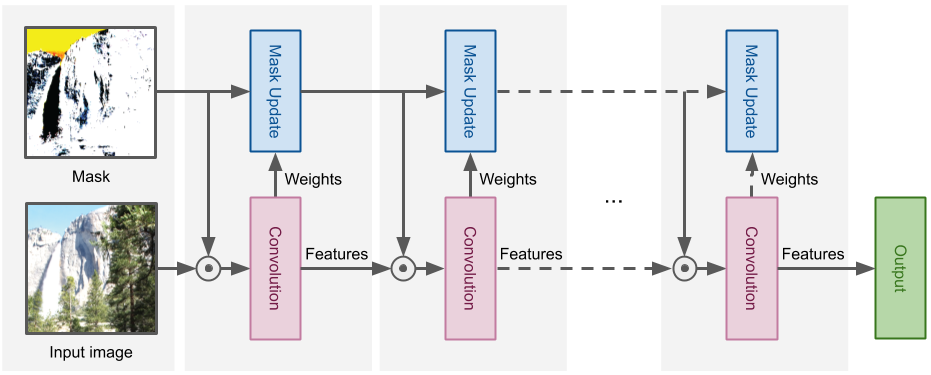}
  \vspace{-0.20in}
  \caption{Illustration of the proposed feature masking mechanism. The features at each layer are multiplied with the corresponding mask before going through the convolution process. The masks at each layer are obtained by updating the masks at the previous layer using Eq.~\ref{eqn:03}.}
  \label{fig:feature_masking}
  \vspace{-0.15in}
\end{figure}

We address this problem by proposing a feature masking mechanism (Figure~\ref{fig:feature_masking}) that reduces the magnitude of the features generated from the invalid content (saturated areas). We do this by multiplying the feature maps in each layer by a soft mask, as follows:

\vspace{-9pt}
\begin{equation}
\label{eqn:01}
  Z_{l} = X_{l} \odot M_{l},
\end{equation}
\vspace{-10pt}

\noindent where $ X_{l} \in \mathbb{R}^{H \times W \times C} $ is the feature map of layer $l$ with height $H$, width $W$, and $C$ channels. $ M_{l} \in [0,1]^{H \times W \times C} $ is the mask for layer $l$ and has values in the range $ [0, 1]$. The value of one indicates that the features are computed from valid input pixels, while zero is assigned to the features that are computed from invalid pixels. \mycolor{Here}, $l = 1$ refers to the input layer and, thus, $X_{l=1}$ is the input LDR image. Similarly, $M_{l=1}$ is the input mask  $M = \beta(T)$. \mycolor{Note that, since our masks are soft, weak signals in the saturated areas are not discarded using this strategy. In fact, by suppressing the invalid pixels, these weak signals can propagate through the network more effectively.} 

\begin{figure}
  \includegraphics[width=1.0 \linewidth]{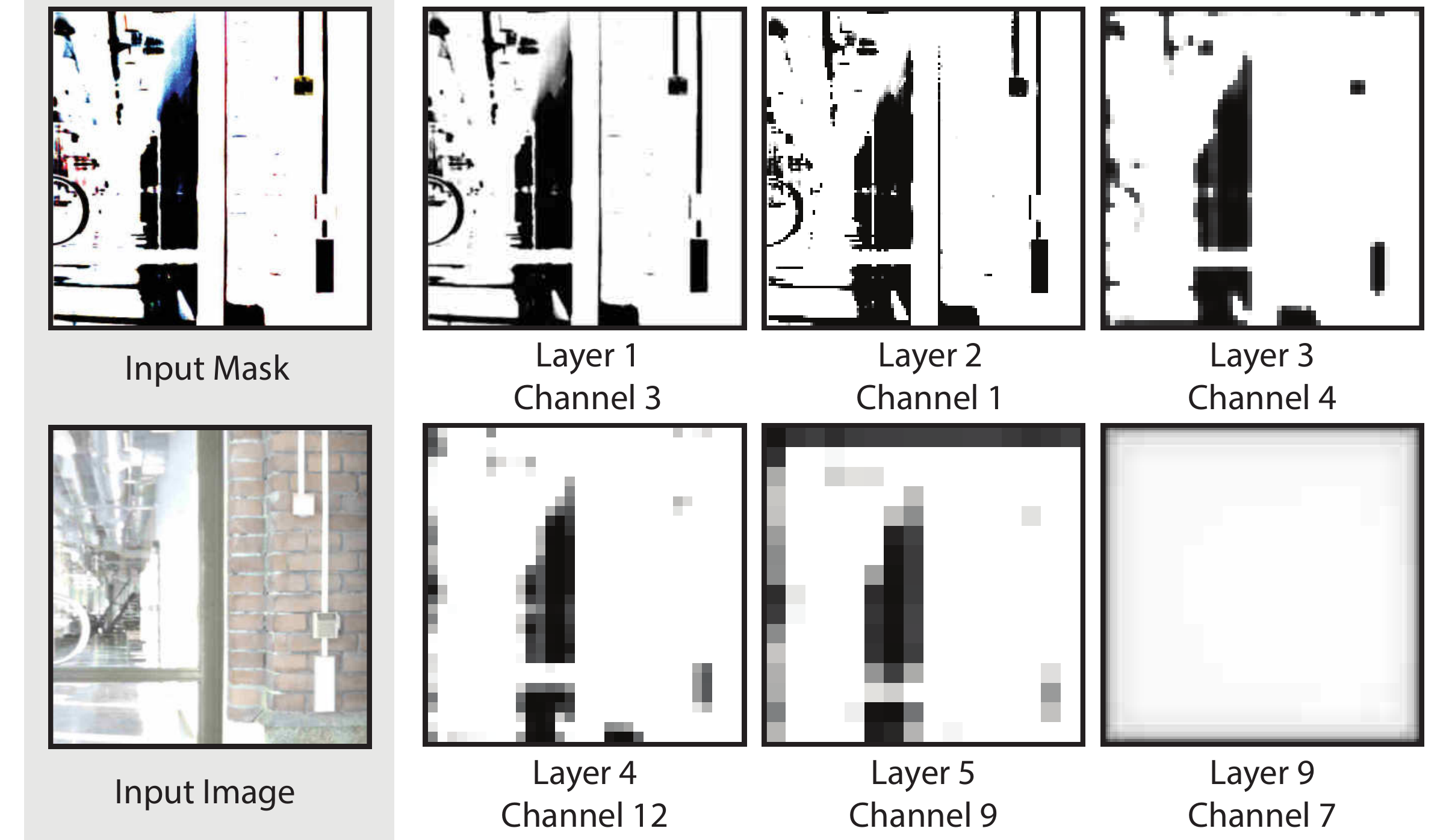}
  \vspace{-0.3in}
  \caption{\mycolor{On the left, we show the input image and the corresponding mask. On the right, we visualize a few masks at different layers of the network. Note that, as we move deeper through the network, the masks become blurrier and more uniform. This is expected since the receptive field of the features become larger in the deeper layers.}}
  \label{fig:masks}
  \vspace{-0.15in}
\end{figure}

Once the features of the current layer $l$ are masked, the features in the next layer $X_{l+1}$ are computed as usual:

\vspace{-8pt}
\begin{equation}
\label{eqn:02}
  X_{l+1} = \phi_l(W_{l}*Z_{l} + b_{l}),
\end{equation}
\vspace{-8pt}

\noindent  where $W_{l}$ and $b_{l}$ refer to the weight and bias of the current layer, respectively. Moreover, $\phi_l$ is the activation function and * is the standard convolution operation. 

We compute the masks at each layer by applying the convolutional filter to the masks at the previous layer \mycolor{(See Figure~\ref{fig:masks} for visualization of some of the masks).} The basic idea is that since the features are computed by applying a series of convolutions, the same filters can be used to compute the contribution of the valid pixels in the features. However, since the masks are in the range $[0, 1]$ and measure the percentage of the contributions, the magnitude of the filters is irrelevant. Therefore, we normalize the filter weights before convolving them with the masks as follows:

\vspace{-2pt}
\begin{equation}
\label{eqn:03}
\begin{aligned}
  M_{l+1} = \left(\frac{\vert W_{l} \vert}{\Vert W_{l}\Vert_{1} + \epsilon}  \right) * M_{l},
\end{aligned}
\end{equation}
\vspace{-2pt}

\noindent where $\Vert \cdot \Vert_{1}$ is the $l_1$ function and $\vert \cdot \vert$ is the absolute operator. Here, $\vert W_{l} \vert$ is a $\mathbb{R}^{H \times W \times C}$ tensor and $\Vert W_{l} \Vert_{1}$ is a $\mathbb{R}^{1 \times 1 \times C}$ tensor. To perform the division, we replicate the values of $\Vert W_{l} \Vert_{1}$ to obtain a tensor with the same size as $\vert W_{l} \vert $. The constant $\epsilon$ is a small value to avoid division by 0 ($ 10^{-6}$ in our implementation).

Note that a couple of recent approaches have proposed strategies to overcome similar issues in image inpainting \cite{yu2018free, liu2018image}. Specifically, \citet{liu2018image} propose to modify the convolution process to only apply the filter to the pixels with valid information. Unfortunately, this approach is specially designed for cases with binary masks. However, the masks in our application are soft and, thus, this method is not applicable. \citet{yu2018free} propose to multiply the features at each layer with a soft mask, similar to our feature masking strategy. The key difference is that their mask at each layer is learnable, and it is estimated using a small network from the features in the previous layer. Because of the additional parameters and complexity, training this approach on limited HDR images is difficult. Therefore, this approach is not able to produce high-quality HDR images (see Section~\ref{sec:ablation}).

\subsection{Loss Function}
\label{sec:hybridloss}

The choice of the loss function is critical in each learning system. Our goal is to reconstruct an HDR image by synthesizing plausible textures in the saturated areas. Unfortunately, using only pixel-wise loss functions, as utilized by most previous approaches, the network tends to produce \mycolor{blurry images (Figure ~\ref{fig:loss_comp})}. Inspired by the recent image inpainting approaches \cite{yang2017high, liu2018image, han2019finet}, we train our network using a VGG-based perceptual loss function. Specifically, our loss function is a combination of an HDR reconstruction loss $ {L}_{r} $ and a perceptual loss $ {L}_{p} $, as follows:

\vspace{-6pt}
\begin{equation}
    \label{eqn:fullloss}
      L = {\lambda}_1 {L}_{r} + {\lambda}_2 {L}_{p}
\end{equation}
\vspace{-10pt}

\noindent where $ {\lambda}_1 = 6.0$ and ${\lambda}_2 = 1.0$ in our implementation. 
\paragraph{Reconstruction Loss:} The HDR reconstruction loss is a simple pixel-wise $l_1$ distance between the output and ground truth images in the saturated areas. Since the HDR images could potentially have large values, we define the loss in the logarithmic domain. Given the estimated HDR image $\hat{Y}$ (in the log domain) and the  linear ground truth image $H$, the reconstruction loss is defined as:

\vspace{-8pt}
\begin{equation}
    {L}_{r} = \Vert (1-M) \odot (\hat{Y} - \log(H + 1)) \Vert_{1}.
\end{equation}
\vspace{-8pt}

\noindent The multiplication by $(1-M)$ ensures that the loss is computed in the saturated areas.

\paragraph{Perceptual Loss:} Our perceptual term is a combination of the VGG and style loss functions as follows:

\vspace{-8pt}
\begin{equation}
    \label{eqn:hybridloss}
      {L}_{p} = {\lambda}_3 {L}_{v} + {\lambda}_4 {L}_{s}.
\end{equation}
\vspace{-10pt}

In our implementation, we set ${\lambda}_3 = 1.0$ and  ${\lambda}_4 = 120.0$. The VGG loss function ${L}_{v}$ evaluates how well the features of the reconstructed image match with the features extracted from the ground truth. This allows the model to produce textures that are perceptually similar to the ground truth. This loss term is defined as follows:

\vspace{-5pt}
\begin{equation}
\label{eq:VGG}
    {L}_{v} = \sum_{l} \Vert \phi_{l}(\mathcal{T}(\tilde{H})) - \phi_{l}(\mathcal{T}(H)) \Vert_{1}
\end{equation}
\vspace{-5pt}

\noindent where $\phi_l$ is the feature map extracted from the $l^{\text{th}}$ layer of the VGG network. Moreover, the image $\tilde{H}$ is obtained by combining the information of the ground truth $H$ in the well-exposed regions and the content of the network's output $ \hat{Y} $ in the saturated areas using the mask $M$, as follows:

\vspace{-6pt}
\begin{equation}
    \tilde{H} =  M \odot H + (1-M) \odot \hat{Y}.
\end{equation}
\vspace{-6pt}

We use $\tilde{H}$ in our loss functions to ensure that the supervision is only provided in the saturated areas. Finally, $ \mathcal{T}(\cdot) $ in Eq.~\ref{eq:VGG} is a function that compresses the range to $[0, 1]$. Specifically, we use the differentiable $\mu$-law range compressor:

\vspace{-5pt}
\begin{equation}
\label{eqn:mulaw}
  \mathcal{T}(H) = \frac{\log(1+ \mu H)}{\log(1+ \mu)},
\end{equation}
\vspace{-5pt}

\noindent where $\mu$ is a parameter defining the amount of compression ($\mu = 500$ in our implementation). This is done to ensure that the input to the VGG network is similar to the ones that it has been trained on. 

The style loss in Eq.~\ref{eqn:hybridloss} (${L}_{s}$) captures style and texture by comparing global statistics with a \mycolor{Gram matrix \cite{gatys2015neural}} collected over the entire image. Specifically, the style loss is defined as:

\vspace{-5pt}
\begin{equation}
    {L}_{s} =  \sum_{l} \Vert G_{l}(\mathcal{T}(\tilde{H})) - G_{l}(\mathcal{T}(H)) \Vert_{1},
\end{equation}
\vspace{-5pt}

\noindent where $G_{l}(X)$ is the \mycolor{Gram} matrix of the features in layer $l$ and is defined as follows:

\vspace{-10pt}
\begin{equation}
    G_{l}(X) = \frac{1}{K_l} \phi_{l}(X)^{T} \phi_{l}(X).
\end{equation}
\vspace{-10pt}

\noindent Here, $K_l$ is a normalization factor computed as ${C_lH_lW_l}$. Note that, the feature $\phi_l$ is a matrix of shape $(H_lW_l) \times C_l$ and, thus, the \mycolor{Gram} matrix has a size of $C_l \times C_l$. In our implementation, we use the VGG-19 \cite{simonyan2014very} network and extract features from layers \texttt{pool1}, \texttt{pool2} and \texttt{pool3}. 


\begin{figure}
  \includegraphics[width=0.85 \linewidth]{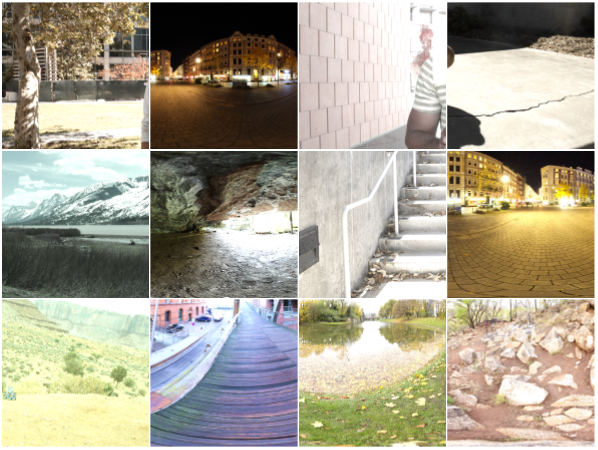}
  \vspace{-0.15in}
  \caption{A few example patches selected by our patch sampling approach. These are challenging examples as the HDR images corresponding to these patches contain complex textures in the saturated areas.}
  \label{fig:patches_selection}
  \vspace{-0.20in}
\end{figure}

\subsection{\mycolor{Inpainting Pre-training}}
\label{sec:train}

Training our system is difficult as large-scale HDR image datasets are currently not available. Existing techniques~\cite{eilertsen2017hdr} overcome this limitation by pre-training their network on simulated HDR images that are created from standard image datasets like the MIT Places~\cite{zhou2014learning}. They then fine-tune their network on real HDR images. Unfortunately, our network is not able to learn to synthesize plausible textures with this strategy (see Figure~\ref{fig:two_stage}), as the saturated areas are typically in the bright and smooth regions.

To address this problem, we propose to pre-train our network on image inpainting tasks. Intuitively, during inpainting, our network leverages a large number of training data to learn an appropriate internal representation that is capable of synthesizing visually pleasing textures. In the HDR fine-tuning stage, the network adapts the learned representation to the HDR domain to be able to synthesize HDR textures. We follow Liu et al.'s approach~\shortcite{liu2018image} and use their loss function and mask generation strategy during pre-training. Note that we still use our feature masking mechanism for pre-training, but the input masks are binary. We fine-tune the network on real HDR images using the loss function, discussed in Section~\ref{sec:hybridloss}. 

One major problem is that the majority of the bright areas in the HDR examples are smooth and textureless. Therefore, during fine-tuning, the network adapts to these types of patches and, as a result, has difficulty producing textured results (see Figure~\ref{fig:two_stage}). In the next section, we discuss our strategy to select textured and challenging patches.

\begin{algorithm}[t]
\caption{Patch Sampling}
\label{alg:patches_selection}
\begin{algorithmic}[1]
\Procedure{PatchMetric}{$H$, $M$}
    \State $H$: HDR image, $M$: Mask
    \State $\sigma_c = 100.0$ \Comment{Bilateral filter color sigma}
    \State $\sigma_s = 10.0$ \Comment{Bilateral filter space sigma}
    \State $I$ = RgbToGray($H$)
    \State L = log($I$ + 1)
    \State B = bilateralFilter($L,\sigma_c,\sigma_s$)
    \Comment{Base Layer}
    \State D = L - B
    \Comment{Detail Layer}
    \State $G_x$ = getGradX($D$)
    \State $G_y$ = getGradY($D$)
    \State G = abs($G_x$) + abs($G_y$)
    \State \Return mean($G \odot (1-M)$)
\EndProcedure
\end{algorithmic}
\end{algorithm}

\begin{figure}
\centering
\vspace{-0.20in}
\includegraphics[width=\linewidth]{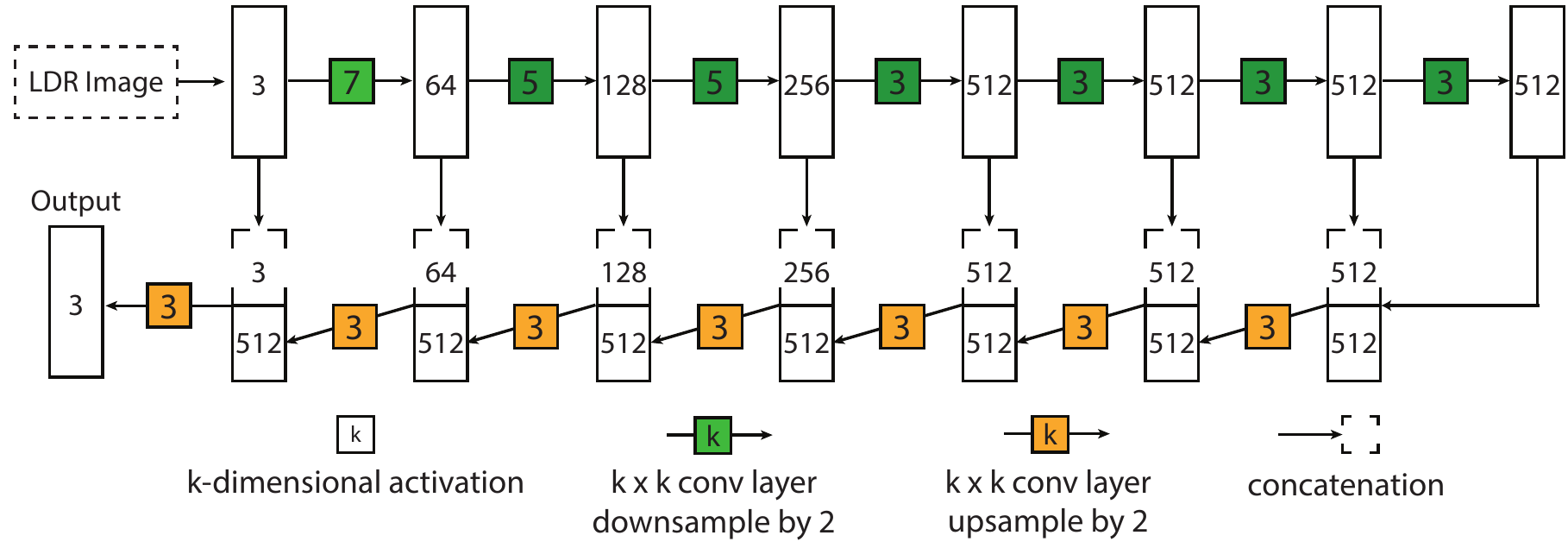}
\vspace{-0.30in}
\caption{The proposed network architecture. The model takes as input the RGB LDR image and outputs an HDR image. We use a feature masking mechanism in all the convolutional layers. }
\vspace{-0.3in}
\label{fig:architecture}
\end{figure}

\begin{figure*}
  \includegraphics[width=\linewidth]{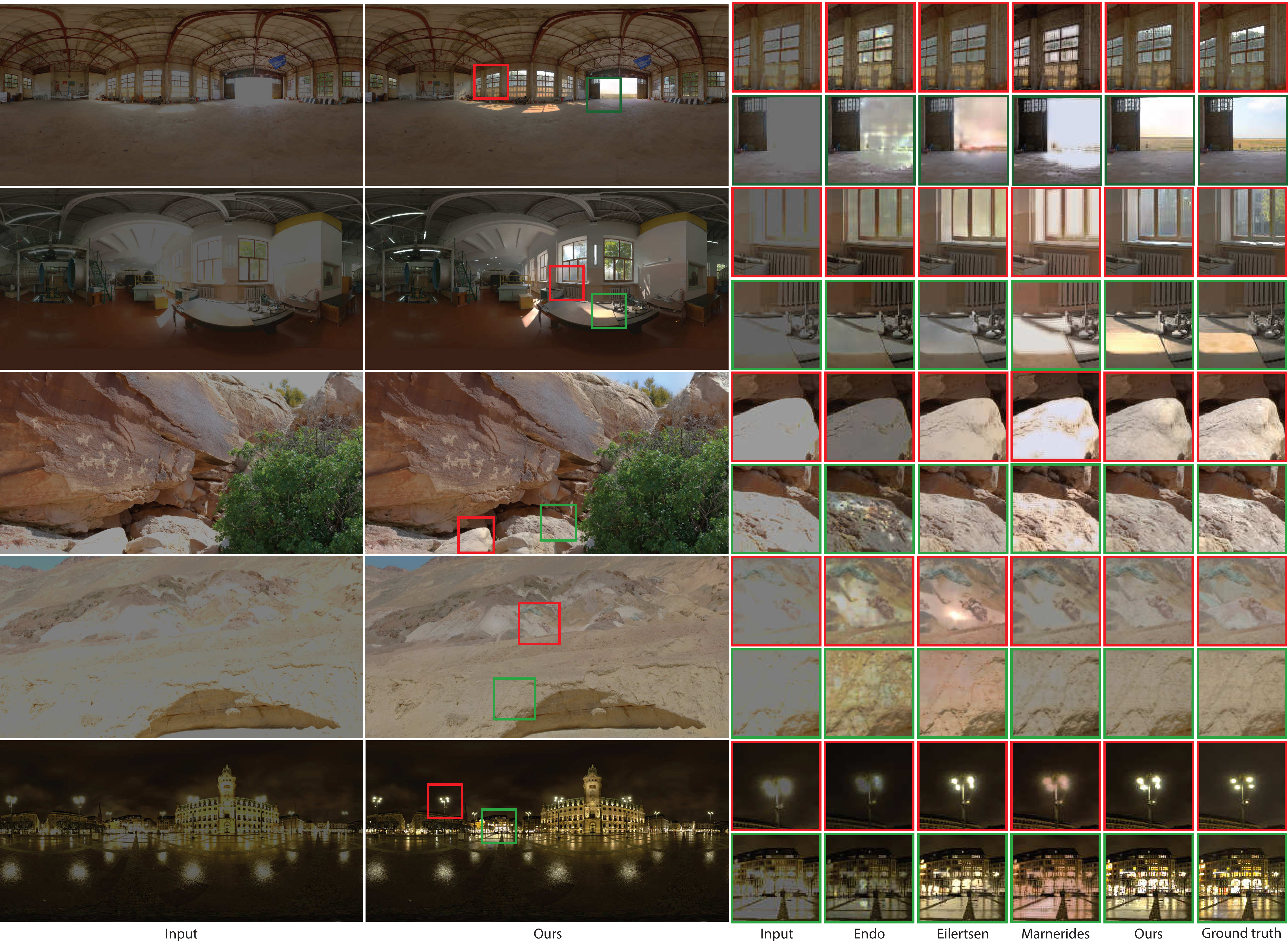}
  \vspace{-0.30in}
  \caption{We compare our method against state-of-the-art approaches of \citet{endo2017deep}, \citet{eilertsen2017hdr}, and \citet{marnerides2018expandnet} on a diverse set of synthetic scenes. Our method is able to synthesize textures in the saturated areas better than the other approaches (rows one to four), while producing results with similar or better quality in the bright highlights (fifth row).}
  \vspace{-0.20in}
  \label{fig:comparison}
\end{figure*}

\subsection{Patch Sampling}
\label{sec:patch_sampling}

Our goal is to select the patches that contain texture in the saturated areas. We perform this by first computing a score for each patch and then choosing the patches with a high score. The main challenge here is finding a good metric that properly detects the textured patches. One way to do this is to compute the average of the gradient magnitude in the saturated regions. However, since our images are in HDR and can have large values, this approach can detect a smooth region with bright highlights as textured.

To avoid this issue, we propose to first decompose the HDR image into base and detail layers using a bilateral filter~\cite{durand2002fast}. We use the average of the gradients (Sobel operator) of the detail layer in the saturated areas as our metric to detect the textured patches. We consider all the patches with a mean gradient above a certain threshold ($0.85$ in our implementation) as textured, and the rest are classified as smooth. Since the detail layer only contains variations around the base layer, this metric can effectively measure the amount of textures in an HDR patch. Figure~\ref{fig:patches_selection} shows example of patches selected using this metric. As shown in Figure~\ref{fig:two_stage}, this simple patch sampling approach is essential for synthesizing HDR images with sharp and artifact-free details in the saturated areas. The summary of our patch selection strategy is listed in Algorithm~\ref{alg:patches_selection}.


\section{Implementation}
\label{sec:architecture}

\paragraph{Architecture}
We use a network with U-Net architecture~\cite{ronneberger2015u}, as shown in Figure~\ref{fig:architecture}. We use the feature masking strategy in all the convolutional layers and up-sample the features in the decoder using nearest neighbor. All the encoder layers use Leaky ReLU activation function~\cite{maas2013rectifier}. On the other hand, we use ReLU~\cite{nair2010rectified} in all the decoder layers, with the exception of the last one, which has a linear activation function. We use skip connections between all the encoder layers and their corresponding decoder layers.

\paragraph{Dataset}
We use different datasets for each training step. For the image inpainting step, we use the MIT Places~\cite{zhou2014learning} dataset with the original train, test, and validation splits. We choose Places for this step because it contains a large number of scenes ($\sim 2.5M$ images) with diverse textures. We use the method of \citet{liu2018image} to generate masks of random streaks and holes of arbitrary shapes and sizes. On the other hand, for the HDR fine-tuning step, we collect approximately 2,000 HDR images from 735 HDR images and 34 HDR videos. From each HDR image, we extract $250$ random patches of size $512 \times 512$ and generate the input LDR patches following the approach by Eilertsen et al.~\shortcite{eilertsen2017hdr}. We then select a subset of these patches using our patch selection strategy. We also discard patches with no saturated content, since they do not provide any source of learning to the network. Our final training dataset is a set of 100K input and corresponding ground truth patches.

\vspace{-0.15pt}
\paragraph{Training}
We initialize our network using the Xavier approach \cite{glorot2010understanding} and train it on image inpainting task until convergence. We then fine-tune the network on HDR reconstruction. We train the network with a learning rate of $ 2 \times 10^{-4} $ in both stages. However, during the second stage, we reduce the learning rate by a factor of $2.0$ when the optimization plateaus. The training process is performed until convergence. Both inpainting and HDR fine-tuning stages are optimized using Adam \cite{kingma2014adam} with the default parameters $\beta_1 = 0.9$ and $\beta_2 = 0.999$ and mini-batch size of 4. The entire training takes approximately 11 days on a machine with an Intel Core i7, 16GB of memory, and an Nvidia GTX 1080 Ti GPU. 

\begin{table}
\hspace*{-2cm}
\caption{Numerical comparison in terms of mean square error (MSE) and HDR-VDP-2 \cite{mantiuk2011hdr} against existing learning-based single image HDR reconstruction approaches.}
\vspace{-0.1in}
\begin{tabular}{lll}\hline
\textbf{Method}    & \multicolumn{1}{c}{\textbf{MSE}}   & \multicolumn{1}{c}{\textbf{HDR-VDP-2}}        \\\hline
\citet{endo2017deep}  &  \multicolumn{1}{c}{0.0390} &  \multicolumn{1}{c}{55.67}\\
\citet{eilertsen2017hdr}  &  \multicolumn{1}{c}{0.0387} &  \multicolumn{1}{c}{59.11} \\
\citet{marnerides2018expandnet}  & \multicolumn{1}{c}{0.0474}  &  \multicolumn{1}{c}{54.31} \\
Ours  & \multicolumn{1}{c}{\textbf{0.0356}}  & \multicolumn{1}{c}{\textbf{63.18}} \\\hline
\end{tabular}
\label{table:existing}
\vspace{-0.2in}
\end{table}

\section{Results}
\label{sec:results}
We implement our network in PyTorch \cite{paszke2017automatic}, but write the data pre-processing, data augmentation, and patch sampling code in C++.  We implement the feature masking mechanism using the existing standard convolutional layer in PyTorch. We compare our approach against three existing learning-based single image HDR reconstruction approaches of \citet{endo2017deep}, \citet{eilertsen2017hdr}, and \citet{marnerides2018expandnet}. We use the source code provided by the authors to generate the results for all the other approaches.

\subsection{Synthetic Images}
We begin by quantitatively comparing our approach against the other methods in terms of mean squared error (MSE) and HDR-VDP-2 \cite{mantiuk2011hdr} in Table~\ref{table:existing}. \mycolor{The errors are computed on a test set of 75 randomly selected HDR images, with resolutions ranging from $1024\times768$ to $2084\times2844$. We generate the input LDR images using various camera curves and exposures, similar to the approach by \citet{eilertsen2017hdr}. We compute the MSE values on the gamma corrected images and HDR-VDP-2 scores are obtained on the linear HDR images.}
As seen, our method produces significantly better results, which demonstrate the ability of our network to accurately recover the full range of \mycolor{luminance}.

Next, we compare our approach against the other methods on five challenging scenes in Figure~\ref{fig:comparison}. Overall other approaches are not able to synthesize texture and produce results with blurriness, discoloration, and checkerboard artifacts. However, our approach can effectively utilize the information in the non-saturated color channels and the contextual information to synthesize visually pleasing textures. It is worth noting that although our approach has been trained using a perceptual loss, it can still properly recover the bright highlights. For example, our results in Figure~\ref{fig:comparison} (fifth row) are similar to \citet{eilertsen2017hdr} and better than \citet{endo2017deep} and \citet{marnerides2018expandnet}.

\begin{figure}
  \includegraphics[width=0.85 \linewidth]{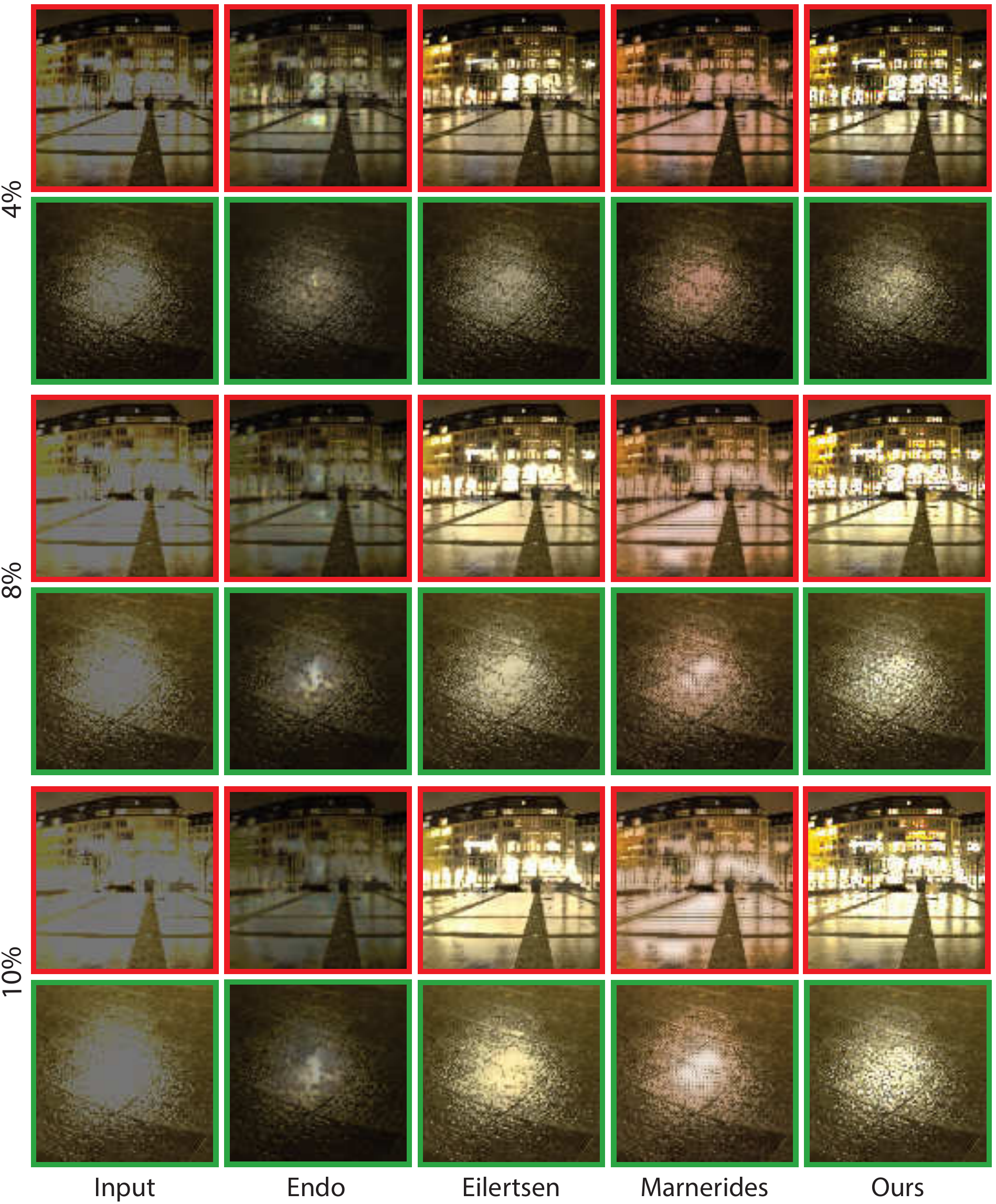}
  \vspace{-0.15in}
  \caption{We compare the performance of the proposed method against previous methods for various amounts of saturated areas. The numbers indicate the percentage of the total number of pixels that are saturated in the input. Although our method slightly degrades as the saturation increases, we consistently present better results than the previous methods.}
  \label{fig:ligh_saturation}
  \vspace{-0.20in}
\end{figure}

We also demonstrate that our approach can consistently generate high-quality results on images with different amount of saturated areas in Figure~\ref{fig:ligh_saturation}. As can be seen, the results of all the other approaches degrade quickly by increasing the percentage of the saturated pixels in the input LDR image. On the other hand, our approach is able to produce high-quality results with sharp details and bright highlights in all the cases.

\begin{figure*}
  \includegraphics[width=\linewidth]{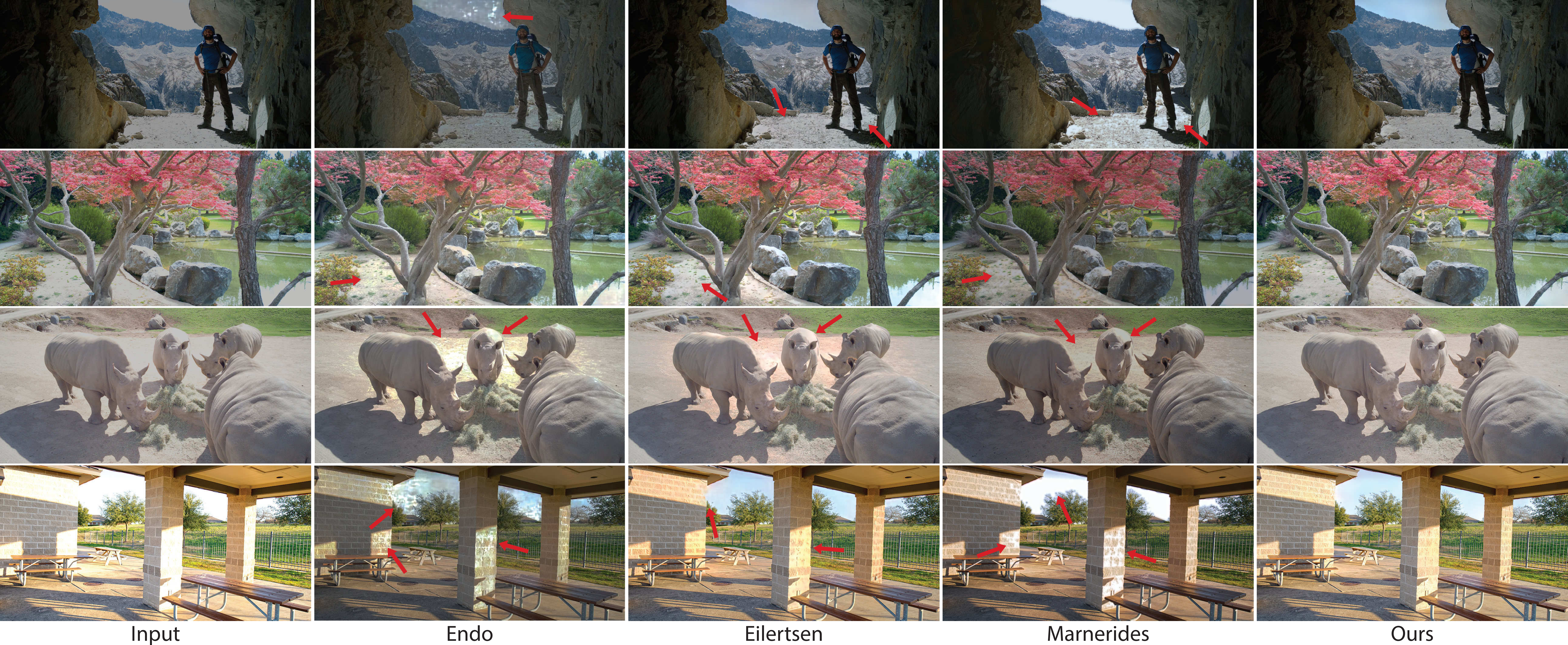}
  \vspace{-0.30in}
  \caption{Comparison against state-of-the-art approaches on images captured by standard cameras. Zoom in to the electronic version to see the differences.}
  \vspace{-0.2in}
  \label{fig:real_world}
\end{figure*}

\subsection{Real Images}
We show the generality of our approach by producing results on a set of real images, captured with standard cameras, in Figure~\ref{fig:real_world}. Specifically, the top three images are from Google HDR+ dataset~\cite{hasinoff2016burst}, captured with a variety of smartphones, such as Nexus 5/6/5X/6P, Pixel, and Pixel XL. The image in the last row is captured by a Canon 5D Mark IV camera. All the other approaches are not able to properly reconstruct the saturated regions, producing results with discoloration and blurriness, as indicated by the arrows. On the other hand, our method is able to properly increase the dynamic range by synthesizing realistic textures.

\begin{table}
\hspace*{-2cm}
\caption{\mycolor{We evaluate the effectiveness of our masking and pre-training strategies by comparing against other alternatives in terms of MSE and HDR-VDP-2 \cite{mantiuk2011hdr}. Here, SConv, GConv, IMask, and FMask refer to standard convolution, gated convolution~\cite{yu2018free}, only masking the input image, and our full feature masking approach, respectively. Moreover, Inp. pre-training and HDR pre-training  correspond to our proposed pre-training on inpainting and HDR reconstruction tasks, respectively. }}
\vspace{-0.1in}
\begin{tabular}{lll}\hline
\textbf{Method (Masking + Pre-training)}    & \multicolumn{1}{c}{\textbf{MSE}}   & \multicolumn{1}{c}{\textbf{HDR-VDP-2}}        \\\hline
\mycolor{SConv + HDR pre-training}  & \multicolumn{1}{c}{\mycolor{0.0402}} &  \multicolumn{1}{c}{\mycolor{58.43}} \\
\mycolor{SConv + Inp. pre-training}  & \multicolumn{1}{c}{\mycolor{0.0374}} &  \multicolumn{1}{c}{\mycolor{60.03}} \\
GConv + HDR pre-training  & \multicolumn{1}{c}{0.0398} &  \multicolumn{1}{c}{53.32} \\
GConv + Inp. pre-training  & \multicolumn{1}{c}{0.1017}  &  \multicolumn{1}{c}{43.13}\\
\mycolor{IMask + HDR pre-training}  & \multicolumn{1}{c}{\mycolor{0.0398}} &  \multicolumn{1}{c}{\mycolor{58.39}} \\
\mycolor{IMask + Inp. pre-training}  & \multicolumn{1}{c}{\mycolor{0.0369}} &  \multicolumn{1}{c}{\mycolor{61.27}} \\
FMask + HDR pre-training  & \multicolumn{1}{c}{\mycolor{0.0393}} & \multicolumn{1}{c}{\mycolor{58.81}}  \\
FMask + Inp. pre-training (Ours)  & \multicolumn{1}{c}{\textbf{0.0356}}  & \multicolumn{1}{c}{\textbf{63.18}} \\\hline
\end{tabular}
\label{table:ablations}
\vspace{-0.20in}
\end{table}

\subsection{Ablation Studies}
\label{sec:ablation}

\begin{figure}
  \includegraphics[width= \linewidth]{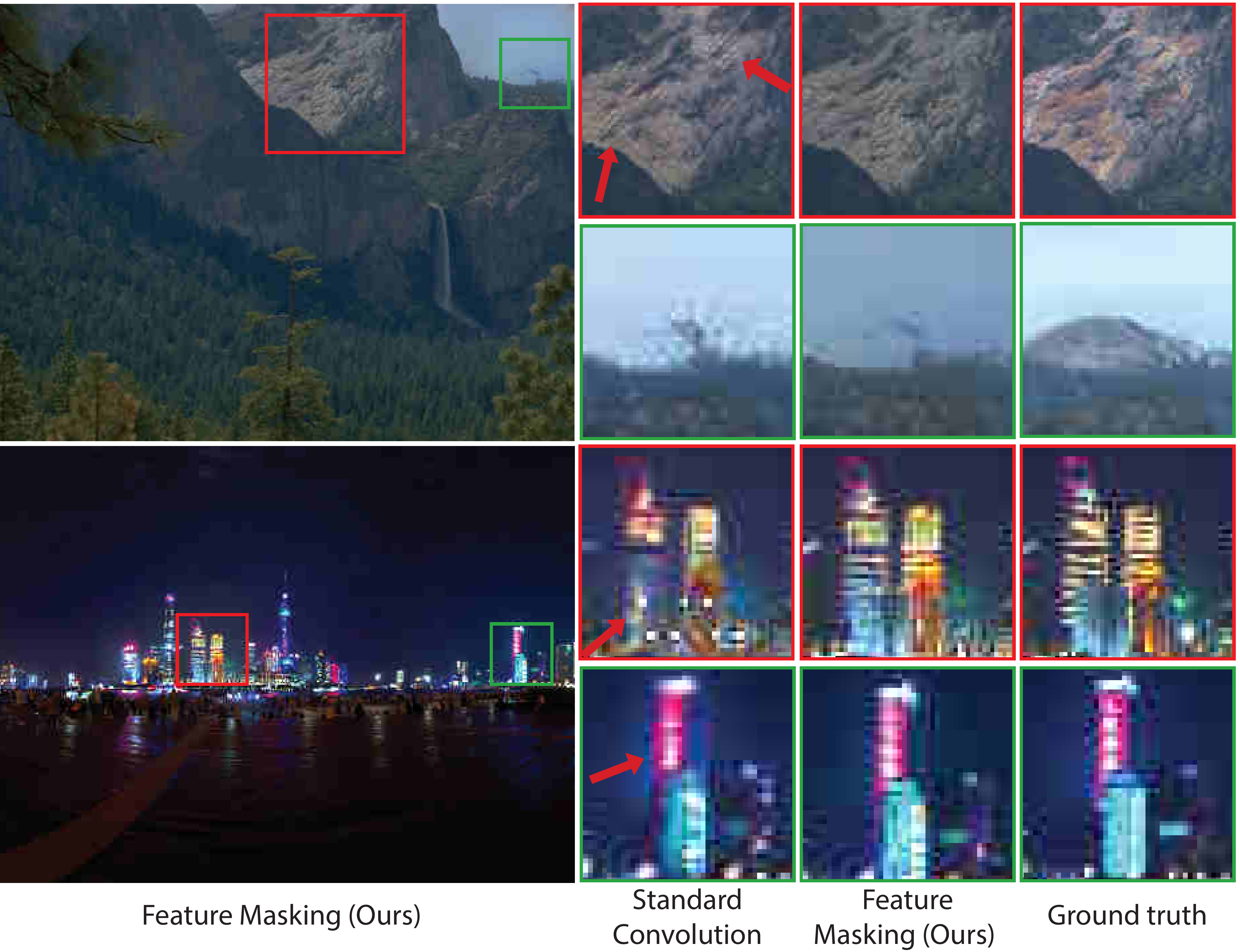}
  \vspace{-0.32in}
  \caption{\mycolor{In regions with both saturated and well-exposed content (boundaries of sky and mountain and bright building lights), the response of the invalid saturated areas in standard convolution dominates the feature maps. Therefore, the network cannot properly utilize the content of the valid regions, introducing high frequency checkerboard artifacts (top row) and blurriness and halo (bottom row). Our approach suppresses the features from the saturated content and allows the network to synthesize the image using the well-exposed information.} }
  \label{fig:masking_std}
  \vspace{-0.25in}
\end{figure}


\vspace{-0.2in}

\begin{figure}
  \includegraphics[width=\linewidth]{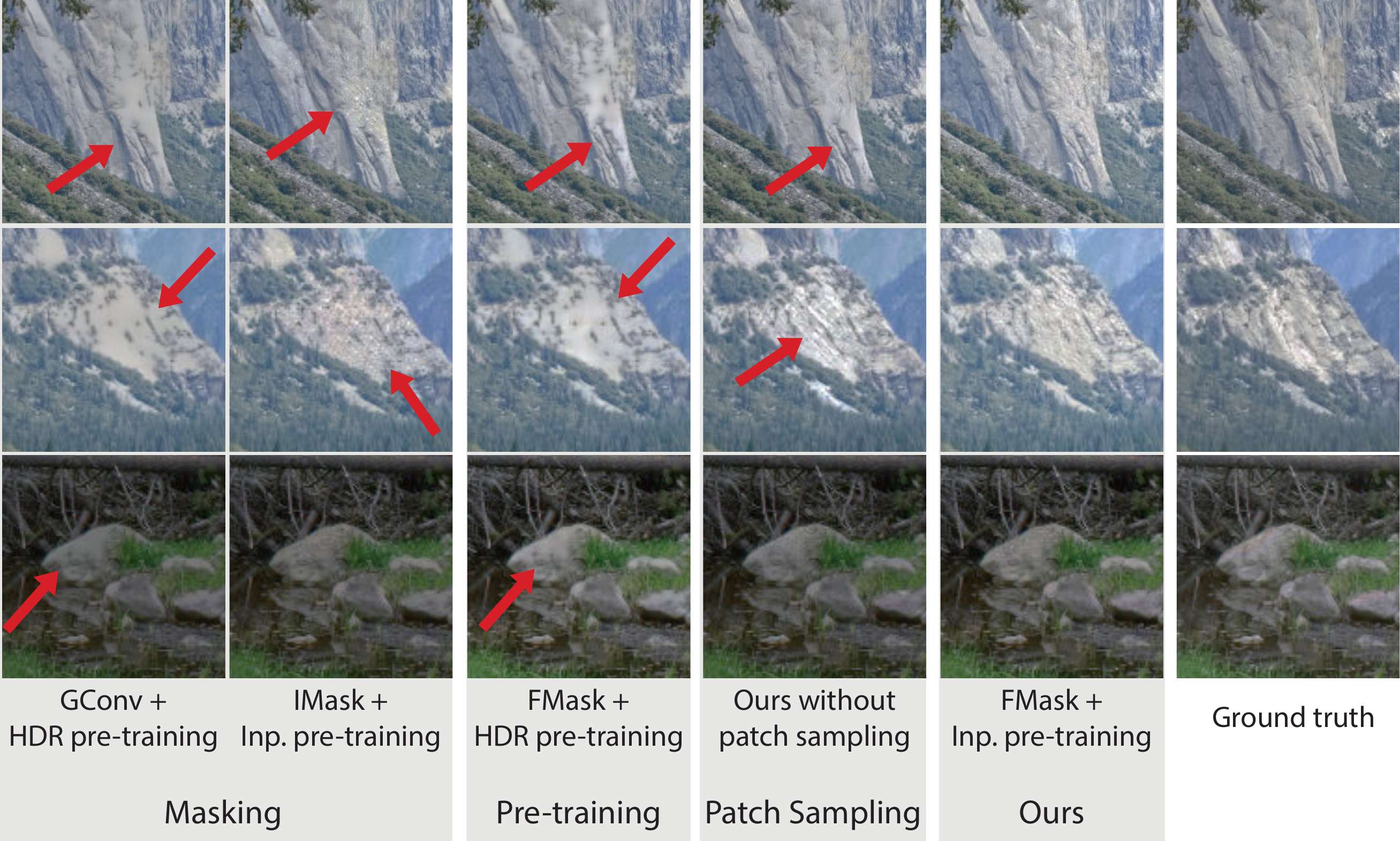}
  \vspace{-20pt}
  \caption{\mycolor{From left to right, we compare our method against two other masking strategies as well as a pre-training method, and evaluate the effect of patch sampling. Here, GConv, IMask, and FMask refer to gated convolution~\cite{yu2018free}, only masking the input image, and our full feature masking method, respectively. Moreover, Inp. pre-training refers to our proposed pre-training on inpainting task.}}
  \label{fig:two_stage}
  \vspace{-0.20in}
\end{figure}

\mycolor{ \paragraph{Inpainting Pre-training.} We begin studying the effect of the proposed inpainting pre-training step by comparing it against the commonly-used synthetic HDR pre-training in Table~\ref{table:ablations} and Figure~\ref{fig:two_stage}. As seen, our pre-training (``FMask + Inp. pre-training (Ours)'') performs better than HDR pre-training (``FMask + HDR pre-training'') both numerically and visually. Specifically, as shown in Figure~\ref{fig:two_stage}, our network using inpainting pre-training is able to learn better features and synthesizes sharp textures in the saturated areas.}

\vspace{-0.2in}

\mycolor{\paragraph{Feature Masking.} Here, we compare our feature masking strategy against several other approaches in Table~\ref{table:ablations}. Specifically, we compare our method against standard convolution (SConv), gated convolution~\cite{yu2018free} (GConv), and the simpler version of our masking strategy where the mask is only applied to the input (IMask). For completeness, we include the result of each method with both inpainting and HDR pre-training. As seen, our masking strategy is considerably better than the other methods. It is worth noting that unlike other methods, the performance of gated convolution with inpainting pre-training is worse than HDR pre-training. This is mainly because gated convolution estimates the masks at each layer using a separate set of networks which become unstable after transitioning from inpainting pre-training to HDR fine-tuning.}

\mycolor{We also visually compare our feature masking method against standard convolution in Figure~\ref{fig:masking_std}. Standard convolution produces results with checkerboard artifacts (top) and halo and blurriness (bottom), while our network with feature masking produces considerably better results. Moreover, we visually compare our approach against other masking strategies in Figure~\ref{fig:two_stage}. Note that, for each masking strategy, we only show the combination of masking and pre-training that produces the best numerical results in Table~\ref{table:ablations}, i.e., gated convolution (GConv) with HDR pre-training and input masking (IMask) with inpainting pre-training. Gated convolution is not able to produce high frequency textures in the saturated areas. Input masking performs reasonably well, but still introduces noticeable artifacts. Our feature masking method, however, is able to synthesize visually pleasing textures.}

\paragraph{Patch Sampling.} We show our result without patch sampling (Section~\ref{sec:patch_sampling}) to demonstrate its effectiveness in Figure~\ref{fig:two_stage}. As seen, by training on the textured patches (ours), the network is able to synthesize textures with more details and fewer objectionable artifacts.

\vspace{-0.2in}

\begin{figure}
  \includegraphics[width=\linewidth]{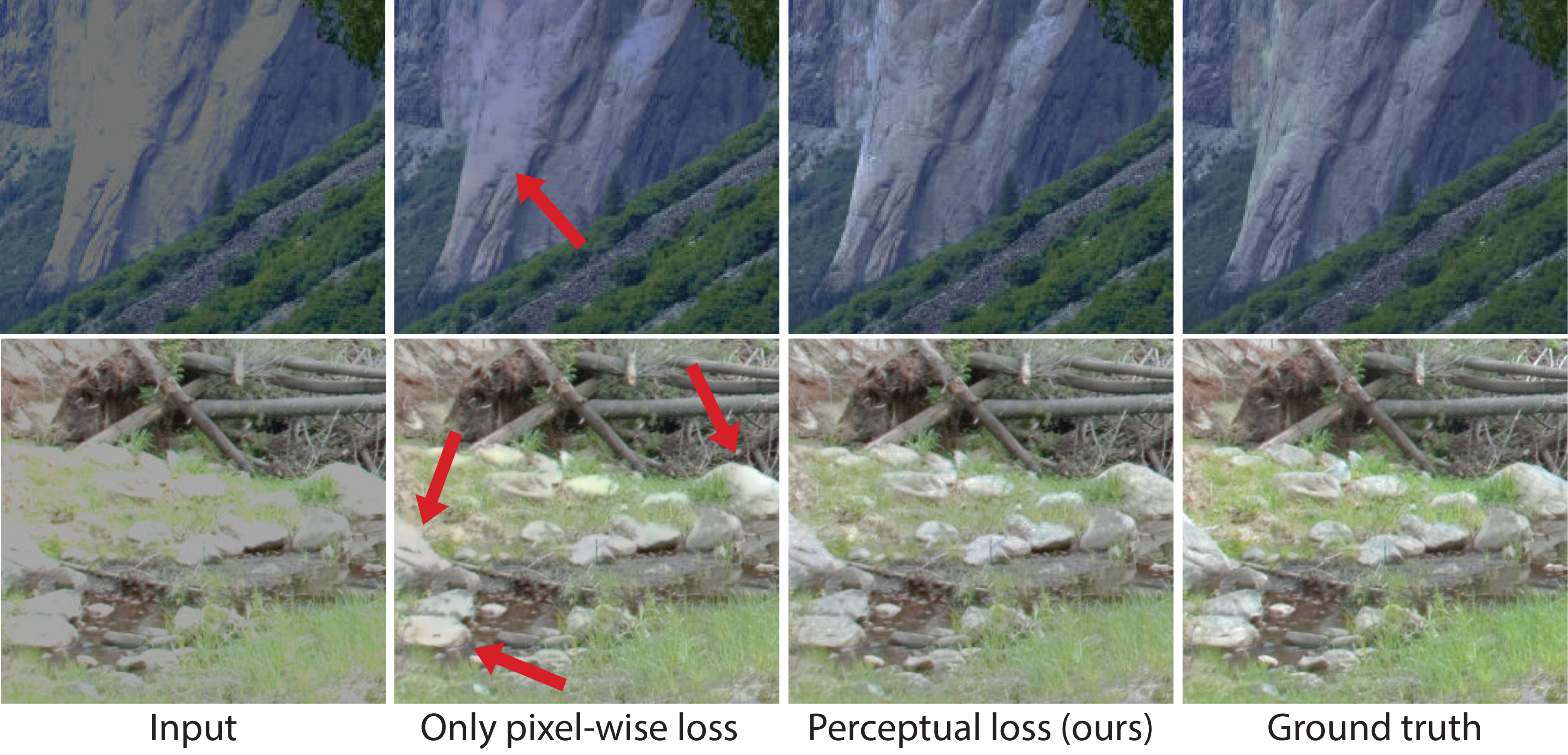}
  \vspace{-0.30in}
  \caption{We compare the results of our network trained with only a pixel-wise loss ($l_1$) and the proposed perceptual loss. Using the perceptual loss function, our network can synthesize visually realistic textures, while the network trained with only a pixel-wise loss produces blurry results.}
  \label{fig:loss_comp}
  \vspace{-0.20in}
\end{figure}

\mycolor{\paragraph{Loss Function.} Finally, we compare the proposed perceptual loss function against a simple pixel-wise ($l_1$) loss. As seen in Figure~\ref{fig:loss_comp}, using only the pixel-wise loss function our network tends to produce blurry images, while the network trained using the proposed perceptual loss function can produce visually realistic textures in the saturated regions.}

\section{Limitations and Future Work}

\begin{figure}
  \includegraphics[width=\linewidth]{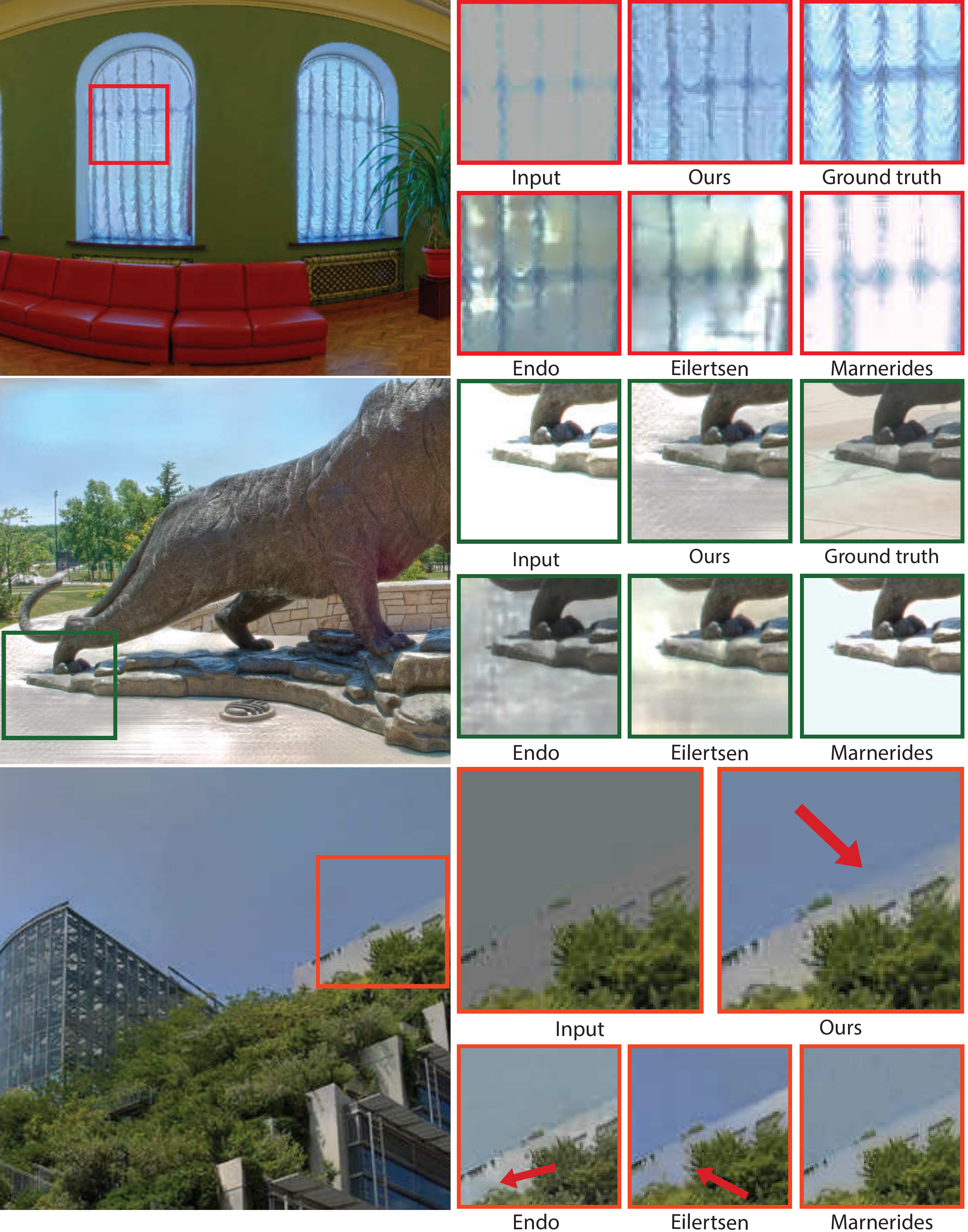}
  \vspace{-20pt}
  \caption{\mycolor{Failure cases of our approach. From top to bottom, our method fails to reconstruct the wrinkles on the curtain, introduces textures that are not in the ground truth, and incorrectly reconstructs the building with sky color. Note that, the top two examples are synthetic, but the bottom one is real for which we do not have access to the ground truth image.}}
  \label{fig:failure_cases}
  \vspace{-0.2in}
\end{figure}

Single image HDR reconstruction is a notoriously challenging problem. Although our method can recover the luminance and hallucinate textures, it is not always able to reconstruct all the details. One of such cases is shown in Figure~\ref{fig:failure_cases} (top), where our approach fails to reconstruct the wrinkles on the curtain. Nevertheless, our result is still better than the other approaches as they overestimate the brightness of the window and produce blurry results. \mycolor{Moreover, as shown in Figure~\ref{fig:failure_cases} (middle), when the input lacks sufficient information about the underlying texture, our method could potentially introduce patterns that do not exist in the ground truth image. Despite that, our result is still comparable to or better than the other approaches. Additionally, in some cases, our method reconstructs the saturated areas with an incorrect color, as shown in Figure~\ref{fig:failure_cases} (bottom). It is worth noting that the network reconstruct the building in blue since trees and skies are usually next to each other in the training data. As seen, other approaches also reconstruct parts of the building in blue color.}

Although our network can be used to reconstruct an HDR video from an LDR video, our result is not temporally stable. This is mainly because we synthesize the content of every frame independently. \mycolor{In the future, it would be interesting to address this problem through temporal regularization~\cite{EMU19}.} Moreover, we would like to experiment with the architecture of the networks to increase the efficiency of our approach and reduce the memory footprint.

\section{Conclusion}
We present a novel learning-based system for single image HDR reconstruction using a convolutional neural network. To alleviate the artifacts caused by conditioning the convolutional layer on the saturated pixels, we propose a feature masking mechanism with an automatic mask updating process. We show that this strategy reduces halo and checkerboard artifacts caused by standard convolutions. Moreover, we propose a perceptual loss function that is designed specifically for the HDR reconstruction application. By minimizing this loss function during training, the network is able to synthesize visually realistic textures in the saturated areas. We further propose to train the system in two stages where we pre-train the network on inpainting before fine-tuning it on HDR generation. To encourage the network to synthesize textures, we propose a sampling strategy to select challenging patches in the HDR examples. Our model can robustly handle saturated areas and can reconstruct high-frequency details in a realistic manner. We show quantitatively and qualitatively that our method outperforms previous methods on both synthetic and real-world images.

\begin{acks}

We thank the reviewers for their constructive comments. M. Santos is funded by the Brazilian agency CNPQ grant 161268/2018-8. T. Ren is partially supported by FACEPE grant APQ-0192- 1.03/14. N. Kalantari is in part funded by a TAMU T3 grant 246451.


\end{acks}

\bibliographystyle{ACM-Reference-Format}
\bibliography{sample-bibliography}

%% file: sample-acmtog-SIGGRAPH-submission.bbl

\begin{thebibliography}{50}


\ifx \showCODEN    \undefined \def \showCODEN     #1{\unskip}     \fi
\ifx \showDOI      \undefined \def \showDOI       #1{#1}\fi
\ifx \showISBNx    \undefined \def \showISBNx     #1{\unskip}     \fi
\ifx \showISBNxiii \undefined \def \showISBNxiii  #1{\unskip}     \fi
\ifx \showISSN     \undefined \def \showISSN      #1{\unskip}     \fi
\ifx \showLCCN     \undefined \def \showLCCN      #1{\unskip}     \fi
\ifx \shownote     \undefined \def \shownote      #1{#1}          \fi
\ifx \showarticletitle \undefined \def \showarticletitle #1{#1}   \fi
\ifx \showURL      \undefined \def \showURL       {\relax}        \fi
\providecommand\bibfield[2]{#2}
\providecommand\bibinfo[2]{#2}
\providecommand\natexlab[1]{#1}
\providecommand\showeprint[2][]{arXiv:#2}

\bibitem[\protect\citeauthoryear{Banterle, Ledda, Debattista, and
  Chalmers}{Banterle et~al\mbox{.}}{2006}]%
        {banterle2006inverse}
\bibfield{author}{\bibinfo{person}{Francesco Banterle},
  \bibinfo{person}{Patrick Ledda}, \bibinfo{person}{Kurt Debattista}, {and}
  \bibinfo{person}{Alan Chalmers}.} \bibinfo{year}{2006}\natexlab{}.
\newblock \showarticletitle{Inverse tone mapping}. In
  \bibinfo{booktitle}{\emph{Proceedings of the 4th International Conference on
  Computer Graphics and Interactive Techniques in Australasia and Southeast
  Asia}}. ACM, \bibinfo{pages}{349--356}.
\newblock


\bibitem[\protect\citeauthoryear{Bau, Strobelt, Peebles, Wulff, Zhou, Zhu, and
  Torralba}{Bau et~al\mbox{.}}{2019}]%
        {Bau:Ganpaint:2019}
\bibfield{author}{\bibinfo{person}{David Bau}, \bibinfo{person}{Hendrik
  Strobelt}, \bibinfo{person}{William Peebles}, \bibinfo{person}{Jonas Wulff},
  \bibinfo{person}{Bolei Zhou}, \bibinfo{person}{Jun{-}Yan Zhu}, {and}
  \bibinfo{person}{Antonio Torralba}.} \bibinfo{year}{2019}\natexlab{}.
\newblock \showarticletitle{Semantic Photo Manipulation with a Generative Image
  Prior}.
\newblock \bibinfo{journal}{\emph{ACM Transactions on Graphics (Proceedings of
  ACM SIGGRAPH)}} \bibinfo{volume}{38}, \bibinfo{number}{4}
  (\bibinfo{year}{2019}).
\newblock


\bibitem[\protect\citeauthoryear{Bist, Cozot, Madec, and Ducloux}{Bist
  et~al\mbox{.}}{2017}]%
        {bist2017tone}
\bibfield{author}{\bibinfo{person}{Cambodge Bist}, \bibinfo{person}{R{\'e}mi
  Cozot}, \bibinfo{person}{G{\'e}rard Madec}, {and} \bibinfo{person}{Xavier
  Ducloux}.} \bibinfo{year}{2017}\natexlab{}.
\newblock \showarticletitle{Tone expansion using lighting style aesthetics}.
\newblock \bibinfo{journal}{\emph{Computers \& Graphics}}  \bibinfo{volume}{62}
  (\bibinfo{year}{2017}), \bibinfo{pages}{77--86}.
\newblock


\bibitem[\protect\citeauthoryear{Debevec}{Debevec}{2005}]%
        {debevec2005median}
\bibfield{author}{\bibinfo{person}{Paul Debevec}.}
  \bibinfo{year}{2005}\natexlab{}.
\newblock \showarticletitle{A median cut algorithm for light probe sampling}.
  In \bibinfo{booktitle}{\emph{ACM SIGGRAPH 2005 Posters}}. ACM,
  \bibinfo{pages}{66}.
\newblock


\bibitem[\protect\citeauthoryear{Debevec and Malik}{Debevec and Malik}{1997}]%
        {debevec1997recovering}
\bibfield{author}{\bibinfo{person}{PE Debevec} {and} \bibinfo{person}{J
  Malik}.} \bibinfo{year}{1997}\natexlab{}.
\newblock \showarticletitle{Recovering high dynamic range images}. In
  \bibinfo{booktitle}{\emph{Proceeding of the SPIE: Image Sensors}},
  Vol.~\bibinfo{volume}{3965}. \bibinfo{pages}{392--401}.
\newblock


\bibitem[\protect\citeauthoryear{Didyk, Mantiuk, Hein, and Seidel}{Didyk
  et~al\mbox{.}}{2008}]%
        {didyk2008enhancement}
\bibfield{author}{\bibinfo{person}{Piotr Didyk}, \bibinfo{person}{Rafal
  Mantiuk}, \bibinfo{person}{Matthias Hein}, {and} \bibinfo{person}{Hans-Peter
  Seidel}.} \bibinfo{year}{2008}\natexlab{}.
\newblock \showarticletitle{Enhancement of bright video features for HDR
  displays}. In \bibinfo{booktitle}{\emph{Computer Graphics Forum}},
  Vol.~\bibinfo{volume}{27}. Wiley Online Library, \bibinfo{pages}{1265--1274}.
\newblock


\bibitem[\protect\citeauthoryear{Dong, Loy, He, and Tang}{Dong
  et~al\mbox{.}}{2015}]%
        {dong2015image}
\bibfield{author}{\bibinfo{person}{Chao Dong}, \bibinfo{person}{Chen~Change
  Loy}, \bibinfo{person}{Kaiming He}, {and} \bibinfo{person}{Xiaoou Tang}.}
  \bibinfo{year}{2015}\natexlab{}.
\newblock \showarticletitle{Image super-resolution using deep convolutional
  networks}.
\newblock \bibinfo{journal}{\emph{IEEE Transactions on Pattern Analysis and
  Machine Intelligence}} \bibinfo{volume}{38}, \bibinfo{number}{2}
  (\bibinfo{year}{2015}), \bibinfo{pages}{295--307}.
\newblock


\bibitem[\protect\citeauthoryear{Durand and Dorsey}{Durand and Dorsey}{2002}]%
        {durand2002fast}
\bibfield{author}{\bibinfo{person}{Fr{\'e}do Durand} {and}
  \bibinfo{person}{Julie Dorsey}.} \bibinfo{year}{2002}\natexlab{}.
\newblock \showarticletitle{Fast bilateral filtering for the display of
  high-dynamic-range images}. In \bibinfo{booktitle}{\emph{Proceedings of the
  29th Annual Conference on Computer Graphics and Interactive Techniques}}.
  \bibinfo{pages}{257--266}.
\newblock


\bibitem[\protect\citeauthoryear{Eilertsen, Kronander, Denes, Mantiuk, and
  Unger}{Eilertsen et~al\mbox{.}}{2017}]%
        {eilertsen2017hdr}
\bibfield{author}{\bibinfo{person}{Gabriel Eilertsen}, \bibinfo{person}{Joel
  Kronander}, \bibinfo{person}{Gyorgy Denes}, \bibinfo{person}{Rafa{\l}~K
  Mantiuk}, {and} \bibinfo{person}{Jonas Unger}.}
  \bibinfo{year}{2017}\natexlab{}.
\newblock \showarticletitle{HDR image reconstruction from a single exposure
  using deep CNNs}.
\newblock \bibinfo{journal}{\emph{ACM Transactions on Graphics (TOG)}}
  \bibinfo{volume}{36}, \bibinfo{number}{6} (\bibinfo{year}{2017}),
  \bibinfo{pages}{178}.
\newblock


\bibitem[\protect\citeauthoryear{Eilertsen, Mantiuk, and Unger}{Eilertsen
  et~al\mbox{.}}{2019}]%
        {EMU19}
\bibfield{author}{\bibinfo{person}{Gabriel Eilertsen}, \bibinfo{person}{Rafa\l
  Mantiuk}, {and} \bibinfo{person}{Jonas Unger}.}
  \bibinfo{year}{2019}\natexlab{}.
\newblock \showarticletitle{Single-frame Regularization for Temporally Stable
  CNNs}. In \bibinfo{booktitle}{\emph{The IEEE Conference on Computer Vision
  and Pattern Recognition (CVPR)}}.
\newblock


\bibitem[\protect\citeauthoryear{Endo, Kanamori, and Mitani}{Endo
  et~al\mbox{.}}{2017}]%
        {endo2017deep}
\bibfield{author}{\bibinfo{person}{Yuki Endo}, \bibinfo{person}{Yoshihiro
  Kanamori}, {and} \bibinfo{person}{Jun Mitani}.}
  \bibinfo{year}{2017}\natexlab{}.
\newblock \showarticletitle{Deep reverse tone mapping.}
\newblock \bibinfo{journal}{\emph{ACM Transactions on Graphics (TOG)}}
  \bibinfo{volume}{36}, \bibinfo{number}{6} (\bibinfo{year}{2017}),
  \bibinfo{pages}{177--1}.
\newblock


\bibitem[\protect\citeauthoryear{Gatys, Ecker, and Bethge}{Gatys
  et~al\mbox{.}}{2015}]%
        {gatys2015neural}
\bibfield{author}{\bibinfo{person}{Leon~A Gatys}, \bibinfo{person}{Alexander~S
  Ecker}, {and} \bibinfo{person}{Matthias Bethge}.}
  \bibinfo{year}{2015}\natexlab{}.
\newblock \showarticletitle{A neural algorithm of artistic style}.
\newblock \bibinfo{journal}{\emph{arXiv preprint arXiv:1508.06576}}
  (\bibinfo{year}{2015}).
\newblock


\bibitem[\protect\citeauthoryear{Gatys, Ecker, and Bethge}{Gatys
  et~al\mbox{.}}{2016}]%
        {gatys2016image}
\bibfield{author}{\bibinfo{person}{Leon~A Gatys}, \bibinfo{person}{Alexander~S
  Ecker}, {and} \bibinfo{person}{Matthias Bethge}.}
  \bibinfo{year}{2016}\natexlab{}.
\newblock \showarticletitle{Image style transfer using convolutional neural
  networks}. In \bibinfo{booktitle}{\emph{Proceedings of the IEEE Conference on
  Computer Vision and Pattern Recognition (CVPR)}}.
  \bibinfo{pages}{2414--2423}.
\newblock


\bibitem[\protect\citeauthoryear{Glorot and Bengio}{Glorot and Bengio}{2010}]%
        {glorot2010understanding}
\bibfield{author}{\bibinfo{person}{Xavier Glorot} {and} \bibinfo{person}{Yoshua
  Bengio}.} \bibinfo{year}{2010}\natexlab{}.
\newblock \showarticletitle{Understanding the difficulty of training deep
  feedforward neural networks}. In \bibinfo{booktitle}{\emph{Proceedings of the
  Thirteenth International Conference on Artificial Intelligence and Statistics
  (AISTATS)}}. \bibinfo{pages}{249--256}.
\newblock


\bibitem[\protect\citeauthoryear{Goodfellow, Pouget-Abadie, Mirza, Xu,
  Warde-Farley, Ozair, Courville, and Bengio}{Goodfellow et~al\mbox{.}}{2014}]%
        {goodfellow2014generative}
\bibfield{author}{\bibinfo{person}{Ian Goodfellow}, \bibinfo{person}{Jean
  Pouget-Abadie}, \bibinfo{person}{Mehdi Mirza}, \bibinfo{person}{Bing Xu},
  \bibinfo{person}{David Warde-Farley}, \bibinfo{person}{Sherjil Ozair},
  \bibinfo{person}{Aaron Courville}, {and} \bibinfo{person}{Yoshua Bengio}.}
  \bibinfo{year}{2014}\natexlab{}.
\newblock \showarticletitle{Generative adversarial nets}. In
  \bibinfo{booktitle}{\emph{Advances in Neural Information Processing Systems
  (NeurIPS)}}. \bibinfo{pages}{2672--2680}.
\newblock


\bibitem[\protect\citeauthoryear{Han, Wu, Huang, Scott, and Davis}{Han
  et~al\mbox{.}}{2019}]%
        {han2019finet}
\bibfield{author}{\bibinfo{person}{Xintong Han}, \bibinfo{person}{Zuxuan Wu},
  \bibinfo{person}{Weilin Huang}, \bibinfo{person}{Matthew~R Scott}, {and}
  \bibinfo{person}{Larry~S Davis}.} \bibinfo{year}{2019}\natexlab{}.
\newblock \showarticletitle{FiNet: Compatible and Diverse Fashion Image
  Inpainting}. In \bibinfo{booktitle}{\emph{Proceedings of the IEEE
  International Conference on Computer Vision (ICCV)}}.
  \bibinfo{pages}{4481--4491}.
\newblock


\bibitem[\protect\citeauthoryear{Hasinoff, Sharlet, Geiss, Adams, Barron,
  Kainz, Chen, and Levoy}{Hasinoff et~al\mbox{.}}{2016}]%
        {hasinoff2016burst}
\bibfield{author}{\bibinfo{person}{Samuel~W Hasinoff}, \bibinfo{person}{Dillon
  Sharlet}, \bibinfo{person}{Ryan Geiss}, \bibinfo{person}{Andrew Adams},
  \bibinfo{person}{Jonathan~T Barron}, \bibinfo{person}{Florian Kainz},
  \bibinfo{person}{Jiawen Chen}, {and} \bibinfo{person}{Marc Levoy}.}
  \bibinfo{year}{2016}\natexlab{}.
\newblock \showarticletitle{Burst photography for high dynamic range and
  low-light imaging on mobile cameras}.
\newblock \bibinfo{journal}{\emph{ACM Transactions on Graphics (TOG)}}
  \bibinfo{volume}{35}, \bibinfo{number}{6} (\bibinfo{year}{2016}),
  \bibinfo{pages}{192}.
\newblock


\bibitem[\protect\citeauthoryear{Hinton and Salakhutdinov}{Hinton and
  Salakhutdinov}{2006}]%
        {hinton2006reducing}
\bibfield{author}{\bibinfo{person}{Geoffrey~E Hinton} {and}
  \bibinfo{person}{Ruslan~R Salakhutdinov}.} \bibinfo{year}{2006}\natexlab{}.
\newblock \showarticletitle{Reducing the dimensionality of data with neural
  networks}.
\newblock \bibinfo{journal}{\emph{Science}} \bibinfo{volume}{313},
  \bibinfo{number}{5786} (\bibinfo{year}{2006}), \bibinfo{pages}{504--507}.
\newblock


\bibitem[\protect\citeauthoryear{Hu, Gallo, Pulli, and Sun}{Hu
  et~al\mbox{.}}{2013}]%
        {hu2013hdr}
\bibfield{author}{\bibinfo{person}{Jun Hu}, \bibinfo{person}{Orazio Gallo},
  \bibinfo{person}{Kari Pulli}, {and} \bibinfo{person}{Xiaobai Sun}.}
  \bibinfo{year}{2013}\natexlab{}.
\newblock \showarticletitle{HDR deghosting: How to deal with saturation?}. In
  \bibinfo{booktitle}{\emph{Proceedings of the IEEE Conference on Computer
  Vision and Pattern Recognition (CVPR)}}. \bibinfo{pages}{1163--1170}.
\newblock


\bibitem[\protect\citeauthoryear{Kalantari and Ramamoorthi}{Kalantari and
  Ramamoorthi}{2017}]%
        {kalantari2017deep}
\bibfield{author}{\bibinfo{person}{Nima~Khademi Kalantari} {and}
  \bibinfo{person}{Ravi Ramamoorthi}.} \bibinfo{year}{2017}\natexlab{}.
\newblock \showarticletitle{Deep high dynamic range imaging of dynamic scenes.}
\newblock \bibinfo{journal}{\emph{ACM Transactions on Graphics (TOG)}}
  \bibinfo{volume}{36}, \bibinfo{number}{4} (\bibinfo{year}{2017}),
  \bibinfo{pages}{144--1}.
\newblock


\bibitem[\protect\citeauthoryear{Kang, Uyttendaele, Winder, and Szeliski}{Kang
  et~al\mbox{.}}{2003}]%
        {kang2003high}
\bibfield{author}{\bibinfo{person}{Sing~Bing Kang}, \bibinfo{person}{Matthew
  Uyttendaele}, \bibinfo{person}{Simon Winder}, {and} \bibinfo{person}{Richard
  Szeliski}.} \bibinfo{year}{2003}\natexlab{}.
\newblock \showarticletitle{High dynamic range video}. In
  \bibinfo{booktitle}{\emph{ACM Transactions on Graphics (TOG)}},
  Vol.~\bibinfo{volume}{22}. ACM, \bibinfo{pages}{319--325}.
\newblock


\bibitem[\protect\citeauthoryear{Kim, Oh, and Kim}{Kim et~al\mbox{.}}{2019}]%
        {kim2019jsi}
\bibfield{author}{\bibinfo{person}{Soo~Ye Kim}, \bibinfo{person}{Jihyong Oh},
  {and} \bibinfo{person}{Munchurl Kim}.} \bibinfo{year}{2019}\natexlab{}.
\newblock \showarticletitle{Jsi-gan: Gan-based joint super-resolution and
  inverse tone-mapping with pixel-wise task-specific filters for UHD HDR
  video}.
\newblock \bibinfo{journal}{\emph{arXiv preprint arXiv:1909.04391}}
  (\bibinfo{year}{2019}).
\newblock


\bibitem[\protect\citeauthoryear{Kingma and Ba}{Kingma and Ba}{2015}]%
        {kingma2014adam}
\bibfield{author}{\bibinfo{person}{Diederick~P Kingma} {and}
  \bibinfo{person}{Jimmy Ba}.} \bibinfo{year}{2015}\natexlab{}.
\newblock \showarticletitle{Adam: A method for stochastic optimization}. In
  \bibinfo{booktitle}{\emph{International Conference on Learning
  Representations (ICLR)}}.
\newblock


\bibitem[\protect\citeauthoryear{Kovaleski and Oliveira}{Kovaleski and
  Oliveira}{2014}]%
        {kovaleski2014high}
\bibfield{author}{\bibinfo{person}{Rafael~P Kovaleski} {and}
  \bibinfo{person}{Manuel~M Oliveira}.} \bibinfo{year}{2014}\natexlab{}.
\newblock \showarticletitle{High-quality reverse tone mapping for a wide range
  of exposures}. In \bibinfo{booktitle}{\emph{2014 27th SIBGRAPI Conference on
  Graphics, Patterns and Images}}. IEEE, \bibinfo{pages}{49--56}.
\newblock


\bibitem[\protect\citeauthoryear{Landis}{Landis}{2002}]%
        {landis2002production}
\bibfield{author}{\bibinfo{person}{Hayden Landis}.}
  \bibinfo{year}{2002}\natexlab{}.
\newblock \showarticletitle{Production-ready global illumination}.
\newblock \bibinfo{journal}{\emph{SIGGRAPH Course Notes}} \bibinfo{volume}{16},
  \bibinfo{number}{2002} (\bibinfo{year}{2002}), \bibinfo{pages}{11}.
\newblock


\bibitem[\protect\citeauthoryear{Lee, An, and Kang}{Lee et~al\mbox{.}}{2018a}]%
        {lee2018deep}
\bibfield{author}{\bibinfo{person}{Siyeong Lee}, \bibinfo{person}{Gwon~Hwan
  An}, {and} \bibinfo{person}{Suk-Ju Kang}.} \bibinfo{year}{2018}\natexlab{a}.
\newblock \showarticletitle{Deep chain hdri: Reconstructing a high dynamic
  range image from a single low dynamic range image}.
\newblock \bibinfo{journal}{\emph{IEEE Access}}  \bibinfo{volume}{6}
  (\bibinfo{year}{2018}), \bibinfo{pages}{49913--49924}.
\newblock


\bibitem[\protect\citeauthoryear{Lee, Hwan~An, and Kang}{Lee
  et~al\mbox{.}}{2018b}]%
        {alee2018deep}
\bibfield{author}{\bibinfo{person}{Siyeong Lee}, \bibinfo{person}{Gwon
  Hwan~An}, {and} \bibinfo{person}{Suk-Ju Kang}.}
  \bibinfo{year}{2018}\natexlab{b}.
\newblock \showarticletitle{Deep Recursive HDRI: Inverse Tone Mapping using
  Generative Adversarial Networks}. In \bibinfo{booktitle}{\emph{Proceedings of
  the European Conference on Computer Vision (ECCV)}}.
  \bibinfo{pages}{596--611}.
\newblock


\bibitem[\protect\citeauthoryear{Liu, Reda, Shih, Wang, Tao, and Catanzaro}{Liu
  et~al\mbox{.}}{2018}]%
        {liu2018image}
\bibfield{author}{\bibinfo{person}{Guilin Liu}, \bibinfo{person}{Fitsum~A
  Reda}, \bibinfo{person}{Kevin~J Shih}, \bibinfo{person}{Ting-Chun Wang},
  \bibinfo{person}{Andrew Tao}, {and} \bibinfo{person}{Bryan Catanzaro}.}
  \bibinfo{year}{2018}\natexlab{}.
\newblock \showarticletitle{Image inpainting for irregular holes using partial
  convolutions}. In \bibinfo{booktitle}{\emph{Proceedings of the European
  Conference on Computer Vision (ECCV)}}. \bibinfo{pages}{85--100}.
\newblock


\bibitem[\protect\citeauthoryear{Luzardo, Aelterman, Luong, Philips, Ochoa, and
  Rousseaux}{Luzardo et~al\mbox{.}}{2018}]%
        {luzardo2018fully}
\bibfield{author}{\bibinfo{person}{Gonzalo Luzardo}, \bibinfo{person}{Jan
  Aelterman}, \bibinfo{person}{Hiep Luong}, \bibinfo{person}{Wilfried Philips},
  \bibinfo{person}{Daniel Ochoa}, {and} \bibinfo{person}{Sven Rousseaux}.}
  \bibinfo{year}{2018}\natexlab{}.
\newblock \showarticletitle{Fully-Automatic Inverse Tone Mapping Preserving the
  Content Creator's Artistic Intentions}. In \bibinfo{booktitle}{\emph{2018
  Picture Coding Symposium (PCS)}}. IEEE, \bibinfo{pages}{199--203}.
\newblock


\bibitem[\protect\citeauthoryear{Maas, Hannun, and Ng}{Maas
  et~al\mbox{.}}{2013}]%
        {maas2013rectifier}
\bibfield{author}{\bibinfo{person}{Andrew~L Maas}, \bibinfo{person}{Awni~Y
  Hannun}, {and} \bibinfo{person}{Andrew~Y Ng}.}
  \bibinfo{year}{2013}\natexlab{}.
\newblock \showarticletitle{Rectifier nonlinearities improve neural network
  acoustic models}. In \bibinfo{booktitle}{\emph{Proceedings of International
  Conference on Machine Learning (ICML)}}, Vol.~\bibinfo{volume}{30}.
  \bibinfo{pages}{3}.
\newblock


\bibitem[\protect\citeauthoryear{Mantiuk, Kim, Rempel, and Heidrich}{Mantiuk
  et~al\mbox{.}}{2011}]%
        {mantiuk2011hdr}
\bibfield{author}{\bibinfo{person}{Rafat Mantiuk}, \bibinfo{person}{Kil~Joong
  Kim}, \bibinfo{person}{Allan~G Rempel}, {and} \bibinfo{person}{Wolfgang
  Heidrich}.} \bibinfo{year}{2011}\natexlab{}.
\newblock \showarticletitle{HDR-VDP-2: A calibrated visual metric for
  visibility and quality predictions in all luminance conditions}.
\newblock \bibinfo{journal}{\emph{ACM Transactions on Graphics (TOG)}}
  \bibinfo{volume}{30}, \bibinfo{number}{4} (\bibinfo{year}{2011}),
  \bibinfo{pages}{40}.
\newblock


\bibitem[\protect\citeauthoryear{Marnerides, Bashford-Rogers, Hatchett, and
  Debattista}{Marnerides et~al\mbox{.}}{2018}]%
        {marnerides2018expandnet}
\bibfield{author}{\bibinfo{person}{Demetris Marnerides},
  \bibinfo{person}{Thomas Bashford-Rogers}, \bibinfo{person}{Jonathan
  Hatchett}, {and} \bibinfo{person}{Kurt Debattista}.}
  \bibinfo{year}{2018}\natexlab{}.
\newblock \showarticletitle{ExpandNet: A deep convolutional neural network for
  high dynamic range expansion from low dynamic range content}. In
  \bibinfo{booktitle}{\emph{Computer Graphics Forum}},
  Vol.~\bibinfo{volume}{37}. Wiley Online Library, \bibinfo{pages}{37--49}.
\newblock


\bibitem[\protect\citeauthoryear{McGuire, Matusik, Pfister, Chen, Hughes, and
  Nayar}{McGuire et~al\mbox{.}}{2007}]%
        {mcguire2007optical}
\bibfield{author}{\bibinfo{person}{Morgan McGuire}, \bibinfo{person}{Wojciech
  Matusik}, \bibinfo{person}{Hanspeter Pfister}, \bibinfo{person}{Billy Chen},
  \bibinfo{person}{John~F Hughes}, {and} \bibinfo{person}{Shree~K Nayar}.}
  \bibinfo{year}{2007}\natexlab{}.
\newblock \showarticletitle{Optical splitting trees for high-precision
  monocular imaging}.
\newblock \bibinfo{journal}{\emph{IEEE Computer Graphics and Applications}}
  \bibinfo{volume}{27}, \bibinfo{number}{2} (\bibinfo{year}{2007}),
  \bibinfo{pages}{32--42}.
\newblock


\bibitem[\protect\citeauthoryear{Nair and Hinton}{Nair and Hinton}{2010}]%
        {nair2010rectified}
\bibfield{author}{\bibinfo{person}{Vinod Nair} {and}
  \bibinfo{person}{Geoffrey~E Hinton}.} \bibinfo{year}{2010}\natexlab{}.
\newblock \showarticletitle{Rectified linear units improve restricted boltzmann
  machines}. In \bibinfo{booktitle}{\emph{Proceedings of the 27th International
  Conference on Machine Learning (ICML)}}. \bibinfo{pages}{807--814}.
\newblock


\bibitem[\protect\citeauthoryear{Ning, Xu, Song, Xie, and Zhang}{Ning
  et~al\mbox{.}}{2018}]%
        {ning2018learning}
\bibfield{author}{\bibinfo{person}{Shiyu Ning}, \bibinfo{person}{Hongteng Xu},
  \bibinfo{person}{Li Song}, \bibinfo{person}{Rong Xie}, {and}
  \bibinfo{person}{Wenjun Zhang}.} \bibinfo{year}{2018}\natexlab{}.
\newblock \showarticletitle{Learning an inverse tone mapping network with a
  generative adversarial regularizer}. In \bibinfo{booktitle}{\emph{2018 IEEE
  International Conference on Acoustics, Speech and Signal Processing
  (ICASSP)}}. IEEE, \bibinfo{pages}{1383--1387}.
\newblock


\bibitem[\protect\citeauthoryear{Oh, Lee, Tai, and Kweon}{Oh
  et~al\mbox{.}}{2014}]%
        {oh2014robust}
\bibfield{author}{\bibinfo{person}{Tae-Hyun Oh}, \bibinfo{person}{Joon-Young
  Lee}, \bibinfo{person}{Yu-Wing Tai}, {and} \bibinfo{person}{In~So Kweon}.}
  \bibinfo{year}{2014}\natexlab{}.
\newblock \showarticletitle{Robust high dynamic range imaging by rank
  minimization}.
\newblock \bibinfo{journal}{\emph{IEEE Transactions on Pattern Analysis and
  Machine Intelligence}} \bibinfo{volume}{37}, \bibinfo{number}{6}
  (\bibinfo{year}{2014}), \bibinfo{pages}{1219--1232}.
\newblock


\bibitem[\protect\citeauthoryear{Paszke, Gross, Massa, Lerer, Bradbury, Chanan,
  Killeen, Lin, Gimelshein, Antiga, et~al\mbox{.}}{Paszke
  et~al\mbox{.}}{2019}]%
        {paszke2017automatic}
\bibfield{author}{\bibinfo{person}{Adam Paszke}, \bibinfo{person}{Sam Gross},
  \bibinfo{person}{Francisco Massa}, \bibinfo{person}{Adam Lerer},
  \bibinfo{person}{James Bradbury}, \bibinfo{person}{Gregory Chanan},
  \bibinfo{person}{Trevor Killeen}, \bibinfo{person}{Zeming Lin},
  \bibinfo{person}{Natalia Gimelshein}, \bibinfo{person}{Luca Antiga},
  {et~al\mbox{.}}} \bibinfo{year}{2019}\natexlab{}.
\newblock \showarticletitle{PyTorch: An imperative style, high-performance deep
  learning library}. In \bibinfo{booktitle}{\emph{Advances in Neural
  Information Processing Systems (NeurIPS)}}. \bibinfo{pages}{8024--8035}.
\newblock


\bibitem[\protect\citeauthoryear{Rempel, Trentacoste, Seetzen, Young, Heidrich,
  Whitehead, and Ward}{Rempel et~al\mbox{.}}{2007}]%
        {rempel2007ldr2hdr}
\bibfield{author}{\bibinfo{person}{Allan~G Rempel}, \bibinfo{person}{Matthew
  Trentacoste}, \bibinfo{person}{Helge Seetzen}, \bibinfo{person}{H~David
  Young}, \bibinfo{person}{Wolfgang Heidrich}, \bibinfo{person}{Lorne
  Whitehead}, {and} \bibinfo{person}{Greg Ward}.}
  \bibinfo{year}{2007}\natexlab{}.
\newblock \showarticletitle{Ldr2hdr: on-the-fly reverse tone mapping of legacy
  video and photographs}. In \bibinfo{booktitle}{\emph{ACM Transactions on
  Graphics (TOG)}}, Vol.~\bibinfo{volume}{26}. ACM, \bibinfo{pages}{39}.
\newblock


\bibitem[\protect\citeauthoryear{Ronneberger, Fischer, and Brox}{Ronneberger
  et~al\mbox{.}}{2015}]%
        {ronneberger2015u}
\bibfield{author}{\bibinfo{person}{Olaf Ronneberger}, \bibinfo{person}{Philipp
  Fischer}, {and} \bibinfo{person}{Thomas Brox}.}
  \bibinfo{year}{2015}\natexlab{}.
\newblock \showarticletitle{U-NET: Convolutional networks for biomedical image
  segmentation}. In \bibinfo{booktitle}{\emph{International Conference on
  Medical Image Computing and Computer-assisted Intervention}}. Springer,
  \bibinfo{pages}{234--241}.
\newblock


\bibitem[\protect\citeauthoryear{Sen, Kalantari, Yaesoubi, Darabi, Goldman, and
  Shechtman}{Sen et~al\mbox{.}}{2012}]%
        {sen2012robust}
\bibfield{author}{\bibinfo{person}{Pradeep Sen}, \bibinfo{person}{Nima~Khademi
  Kalantari}, \bibinfo{person}{Maziar Yaesoubi}, \bibinfo{person}{Soheil
  Darabi}, \bibinfo{person}{Dan~B Goldman}, {and} \bibinfo{person}{Eli
  Shechtman}.} \bibinfo{year}{2012}\natexlab{}.
\newblock \showarticletitle{Robust patch-based HDR reconstruction of dynamic
  scenes.}
\newblock \bibinfo{journal}{\emph{ACM Transactions on Graphics (TOG)}}
  \bibinfo{volume}{31}, \bibinfo{number}{6} (\bibinfo{year}{2012}),
  \bibinfo{pages}{203--1}.
\newblock


\bibitem[\protect\citeauthoryear{Simonyan and Zisserman}{Simonyan and
  Zisserman}{2015}]%
        {simonyan2014very}
\bibfield{author}{\bibinfo{person}{Karen Simonyan} {and}
  \bibinfo{person}{Andrew Zisserman}.} \bibinfo{year}{2015}\natexlab{}.
\newblock \showarticletitle{Very Deep Convolutional Networks for Large-Scale
  Image Recognition}. In \bibinfo{booktitle}{\emph{International Conference on
  Learning Representations (ICLR)}}.
\newblock


\bibitem[\protect\citeauthoryear{Tocci, Kiser, Tocci, and Sen}{Tocci
  et~al\mbox{.}}{2011}]%
        {tocci2011versatile}
\bibfield{author}{\bibinfo{person}{Michael~D Tocci}, \bibinfo{person}{Chris
  Kiser}, \bibinfo{person}{Nora Tocci}, {and} \bibinfo{person}{Pradeep Sen}.}
  \bibinfo{year}{2011}\natexlab{}.
\newblock \showarticletitle{A versatile HDR video production system}. In
  \bibinfo{booktitle}{\emph{ACM Transactions on Graphics (TOG)}},
  Vol.~\bibinfo{volume}{30}. ACM, \bibinfo{pages}{41}.
\newblock


\bibitem[\protect\citeauthoryear{Wang, Wei, Zhou, Guo, and Shum}{Wang
  et~al\mbox{.}}{2007}]%
        {wang2007high}
\bibfield{author}{\bibinfo{person}{Lvdi Wang}, \bibinfo{person}{Li-Yi Wei},
  \bibinfo{person}{Kun Zhou}, \bibinfo{person}{Baining Guo}, {and}
  \bibinfo{person}{Heung-Yeung Shum}.} \bibinfo{year}{2007}\natexlab{}.
\newblock \showarticletitle{High dynamic range image hallucination}. In
  \bibinfo{booktitle}{\emph{Proceedings of the 18th Eurographics Conference on
  Rendering Techniques}}. Eurographics Association, \bibinfo{pages}{321--326}.
\newblock


\bibitem[\protect\citeauthoryear{Wu, Xu, Tai, and Tang}{Wu
  et~al\mbox{.}}{2018}]%
        {wu2018deep}
\bibfield{author}{\bibinfo{person}{Shangzhe Wu}, \bibinfo{person}{Jiarui Xu},
  \bibinfo{person}{Yu-Wing Tai}, {and} \bibinfo{person}{Chi-Keung Tang}.}
  \bibinfo{year}{2018}\natexlab{}.
\newblock \showarticletitle{Deep high dynamic range imaging with large
  foreground motions}. In \bibinfo{booktitle}{\emph{Proceedings of the European
  Conference on Computer Vision (ECCV)}}. \bibinfo{pages}{117--132}.
\newblock


\bibitem[\protect\citeauthoryear{Xu, Ning, Xie, and Song}{Xu
  et~al\mbox{.}}{2019}]%
        {xu2019gan}
\bibfield{author}{\bibinfo{person}{Yucheng Xu}, \bibinfo{person}{Shiyu Ning},
  \bibinfo{person}{Rong Xie}, {and} \bibinfo{person}{Li Song}.}
  \bibinfo{year}{2019}\natexlab{}.
\newblock \showarticletitle{Gan Based Multi-Exposure Inverse Tone Mapping}. In
  \bibinfo{booktitle}{\emph{2019 IEEE International Conference on Image
  Processing (ICIP)}}. IEEE, \bibinfo{pages}{1--5}.
\newblock


\bibitem[\protect\citeauthoryear{Yang, Lu, Lin, Shechtman, Wang, and Li}{Yang
  et~al\mbox{.}}{2017}]%
        {yang2017high}
\bibfield{author}{\bibinfo{person}{Chao Yang}, \bibinfo{person}{Xin Lu},
  \bibinfo{person}{Zhe Lin}, \bibinfo{person}{Eli Shechtman},
  \bibinfo{person}{Oliver Wang}, {and} \bibinfo{person}{Hao Li}.}
  \bibinfo{year}{2017}\natexlab{}.
\newblock \showarticletitle{High-resolution image inpainting using multi-scale
  neural patch synthesis}. In \bibinfo{booktitle}{\emph{Proceedings of the IEEE
  Conference on Computer Vision and Pattern Recognition (CVPR)}}.
  \bibinfo{pages}{6721--6729}.
\newblock


\bibitem[\protect\citeauthoryear{Yang, Xu, Song, Zhang, Wei, and Lau}{Yang
  et~al\mbox{.}}{2018}]%
        {yang2018image}
\bibfield{author}{\bibinfo{person}{Xin Yang}, \bibinfo{person}{Ke Xu},
  \bibinfo{person}{Yibing Song}, \bibinfo{person}{Qiang Zhang},
  \bibinfo{person}{Xiaopeng Wei}, {and} \bibinfo{person}{Rynson~WH Lau}.}
  \bibinfo{year}{2018}\natexlab{}.
\newblock \showarticletitle{Image correction via deep reciprocating HDR
  transformation}. In \bibinfo{booktitle}{\emph{Proceedings of the IEEE
  Conference on Computer Vision and Pattern Recognition (CVPR)}}.
  \bibinfo{pages}{1798--1807}.
\newblock


\bibitem[\protect\citeauthoryear{Yu, Lin, Yang, Shen, Lu, and Huang}{Yu
  et~al\mbox{.}}{2019}]%
        {yu2018free}
\bibfield{author}{\bibinfo{person}{Jiahui Yu}, \bibinfo{person}{Zhe Lin},
  \bibinfo{person}{Jimei Yang}, \bibinfo{person}{Xiaohui Shen},
  \bibinfo{person}{Xin Lu}, {and} \bibinfo{person}{Thomas~S Huang}.}
  \bibinfo{year}{2019}\natexlab{}.
\newblock \showarticletitle{Free-form image inpainting with gated convolution}.
  In \bibinfo{booktitle}{\emph{Proceedings of the IEEE International Conference
  on Computer Vision}}. \bibinfo{pages}{4471--4480}.
\newblock


\bibitem[\protect\citeauthoryear{Zhang, Isola, and Efros}{Zhang
  et~al\mbox{.}}{2016}]%
        {zhang2016colorful}
\bibfield{author}{\bibinfo{person}{Richard Zhang}, \bibinfo{person}{Phillip
  Isola}, {and} \bibinfo{person}{Alexei~A Efros}.}
  \bibinfo{year}{2016}\natexlab{}.
\newblock \showarticletitle{Colorful image colorization}. In
  \bibinfo{booktitle}{\emph{European Conference on Computer Vision (ECCV)}}.
  Springer, \bibinfo{pages}{649--666}.
\newblock


\bibitem[\protect\citeauthoryear{Zhou, Lapedriza, Xiao, Torralba, and
  Oliva}{Zhou et~al\mbox{.}}{2014}]%
        {zhou2014learning}
\bibfield{author}{\bibinfo{person}{Bolei Zhou}, \bibinfo{person}{Agata
  Lapedriza}, \bibinfo{person}{Jianxiong Xiao}, \bibinfo{person}{Antonio
  Torralba}, {and} \bibinfo{person}{Aude Oliva}.}
  \bibinfo{year}{2014}\natexlab{}.
\newblock \showarticletitle{Learning deep features for scene recognition using
  places database}. In \bibinfo{booktitle}{\emph{Advances in Neural Information
  Processing Systems (NeurIPS)}}. \bibinfo{pages}{487--495}.
\newblock


\end{thebibliography}
